\documentclass[letterpaper,twocolumn,10pt]{article}
\usepackage{usenix}

\usepackage{amsmath,amssymb}
\usepackage{blkarray}
\usepackage{booktabs}
\usepackage{graphicx}
\usepackage{multirow}
\usepackage{makecell}
\usepackage{xurl}

\definecolor{bestgreen}{HTML}{C8E6D0}
\newcommand{\best}[1]{\textbf{#1}}

\begin{document}

\date{}

\title{Origin Is All You Need: Provenance-Aware Transformers for Structural Trust-Boundary Separation}
%Origin is All You Need: Provenance-Aware Transformers for Structural Instruction–Data Separation
% for ENforcing Data Provenance and Security

\author{
{\rm Yuxuan Zhang}\\
Texas A\&M University\\
{\small \texttt{yuz516@tamu.edu}}
\and
{\rm Jeff Huang}\\
Texas A\&M University\\
{\small \texttt{jeff@cse.tamu.edu}}
\and
{\rm Guofei Gu}\\
Texas A\&M University\\
{\small \texttt{guofei@cse.tamu.edu}}
}

\maketitle

\begin{abstract}
Indirect prompt injection (IPI) remains a central safety and security challenge for large
language model (LLM) systems because standard transformers lack architectural notion
of source authority. Retrieved documents, user inputs, and system
instructions are all processed through the same undifferentiated attention mechanism, forcing the
model to infer from wording alone what should be obeyed and what should be treated as
data. We propose \emph{Provenance-Aware Transformers}, a provenance-aware defense that makes
application-supplied source labels actionable inside the model. Each input token is
assigned a ring ID encoding its origin, and the model is augmented with origin embeddings,
a learnable origin attention bias, and a learnable origin scale that preserves provenance
under normalization. The resulting architecture enforces a structural boundary between
authoritative and non-authoritative sources during generation. To instantiate this architecture on released pretrained
models, we propose a two-stage fine-tuning pipeline to teach the model origin semantics and task behavior under ring constraints. Evaluation shows that Provenance-Aware Transformers maintain robust
resistance to IPI both in-distribution and out-of-distribution while preserving utility comparable to the base pretrained model.
More broadly, our work shows that exposing provenance as a first-class architectural
signal can shift LLM safety alignment from brittle pattern matching toward explicit trust
separation.
\end{abstract}

% =============================================================================
\section{Introduction}
Large Language Models (LLMs) have become foundational to a wide array of applications from conversational agents and content moderation systems to autonomous planning and tool-augmented workflows~\cite{aws2024llm, ibm2024llm, openai2023gpt4, google2024vertex, meta2024llama2}. Their utility has been further amplified by the rapid adoption of autonomous agent frameworks \cite{claudecode, codex, openclaw, crewai2026, autogen2026}, which enable LLMs to interact with external APIs, execute system commands, and manage persistent memory~\cite{openai2024agents, aws2024agents}. While these developments enable complex, multi-step behaviors, they also significantly expand the attack surface for \emph{Indirect Prompt Injection} (IPI) attacks~\cite{lakera2023visual, Selvi2022, Willison2022, Harang2023}. In these agentic ecosystems, a single malicious instruction from retrieved external resources can subvert intended logic to trigger unauthorized tool execution or data exfiltration across an entire multi-agent workflow\cite{greshake2023not, yi2025benchmarking}. 
% Such vulnerabilities highlight the urgent need for robust defense mechanisms that can ensure the integrity of instruction execution in increasingly autonomous AI systems~\cite{liu2024promptinjection}.

% \begin{figure}[!t]
% \centering
% \includegraphics[width=0.45\textwidth]{figures/introduction.pdf}
% \caption{Illustrative example of IPI attack and Origin-Aware Transformer model defending against IPI attack.}
% \label{fig:intro}
% \end{figure}

\begin{figure}[!t]
\centering
\includegraphics[width=0.35\textwidth]{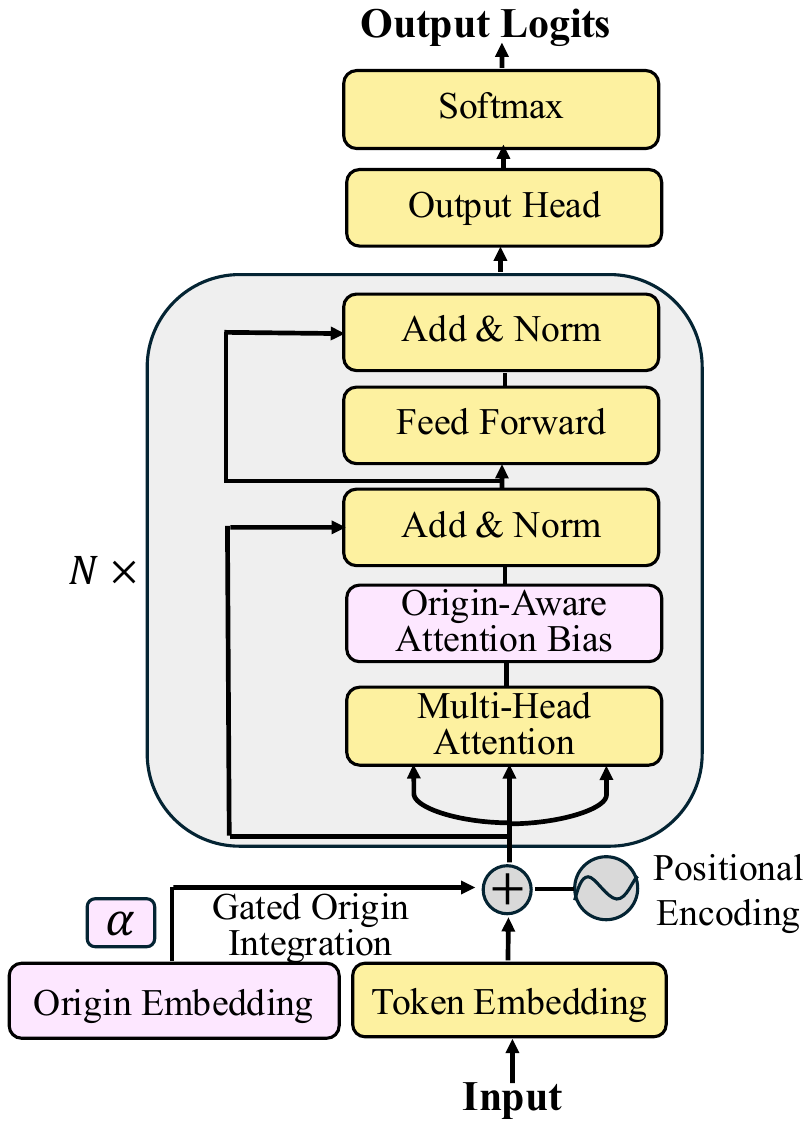}
\caption{Provenance-aware transformer architecture. Yellow blocks denotes the existing components in decoder-only transformers. \textbf{Pink} blocks presents the origin components designed and added to the existing model architecture.}
\label{fig:architecture}
\end{figure}

A growing body of research has explored defenses against (indirect) prompt injection, which could be categorized into \emph{prompt-based},
\emph{detection-based}, and \emph{fine-tuning-based} approaches. Prompt-based defenses
operate at the input level without altering model weights, using techniques such as
instructional reminders~\cite{Wallace2024InstructionHierarchy}, delimiter and
sandwich-style demarcation~\cite{Willison2023, learnprompting2023sandwich}, and other
input-transformation strategies~\cite{Random_Sequnce, Nakajima2022}, but remain
vulnerable to injected content that mimics legitimate instructions. Detection-based
defenses instead treat prompt injection as a classification problem, ranging from
lightweight classifiers~\cite{shi2025promptarmorsimpleeffectiveprompt, promptguard}
to semantic-intent~\cite{wang2025promptsleuth} and attention-based
detectors~\cite{jia2025promptlocate, zhong2025attention} that flag or sanitize contaminated
input before execution, though they operate downstream of generation with no
architectural guarantee against undetected injections. Fine-tuning-based defenses
internalize resistance directly into model weights through preference
optimization~\cite{Chen2024SecAlignDAA} or task-specific training~\cite{Piet2024Jatmo,
chen2025struq}, but rely entirely on the base transformer's content-addressed attention to learn
injection-resistant behavior from examples, without any mechanism that structurally
constrains how untrusted content can influence generation.

% A growing body of research has explored defenses against prompt injection, which can be broadly categorized into \emph{prevention-based} and \emph{detection-based} approaches~\cite{liu2024promptinjection, liu2025datasentinel, zhang2025defense, Chen2024SecAlignDAA, chen2025struq}. Prevention defenses aim to neutralize threats by pre-processing inputs through syntax-level heuristics such as delimiters and sandwich patterns~\cite{Willison2023, learnprompting2023sandwich}, or by utilizing architectural privilege controls~\cite{Debenedetti2025Defeating, Shi2025Progent}. Other prevention methods leverage model-level alignment, such as task-specific fine-tuning~\cite{Piet2024Jatmo} and preference optimization~\cite{chen2025struq, Chen2024SecAlignDAA, Wallace2024InstructionHierarchy}, to ensure the LLM prioritizes privileged instructions. In contrast, detection-based defenses focus on identifying contaminated samples before execution, typically by deploying auxiliary models to flag redirection attempts or anomalous instruction patterns~\cite{liu2025datasentinel, Nakajima2022}. These detectors range from specialized classifiers and filters designed for low-latency screening~\cite{shi2025promptarmorsimpleeffectiveprompt} to generative guardrail models that perform safety-aligned content moderation~\cite{inan2023llama}. By treating detection as an independent verification step, these frameworks attempt to shield the primary LLM from untrusted data.

Despite the breadth of existing defenses, we observe that the root cause of prompt injection remains fundamentally unaddressed: language models possess no architectural mechanism to distinguish \emph{authorized instruction sources} from \emph{non-authoritative data sources}\cite{zverev2025can}, and consequently have no principled structure basis for deciding what to obey based on where content appears. An adversary who embeds a command inside a retrieved document exploits precisely this confusion where the model treats injected text no differently from a legitimate system directive\cite{greshake2023not}.

To address this root cause, we draw on a foundational principle from operating system
security: \emph{origin}, where the source of an operation determines its authorization
rather than its content alone. We adapt this principle to provenance-aware LLM
deployments by assigning each input token a discrete origin label and modifying the
transformer so that provenance directly constrains information flow during generation.
Realizing this idea introduces three challenges. \textbf{First}, labeling tokens alone is
insufficient, the model must be explicitly taught what origin labels mean. Prior
approaches that annotate trust boundaries, such as the sandwich
defense~\cite{Willison2023} and StruQ~\cite{chen2025struq}, remain ineffective because
they still rely on the model's own judgment to honor them. \textbf{Second}, standard
transformer operations actively resist preserving origin information. Normalization
dilutes additive signals as embeddings grow during fine-tuning, and content-addressed
attention provides no native channel for source identity to constrain information flow,
requiring targeted architectural modification. \textbf{Third}, the defense must not
compromise utility on benign inputs, overly aggressive suppression or an insufficiently
diverse training corpus can cause the model to overgeneralize defensive behavior and
degrade ordinary task performance.

Motivated by these challenges, we introduce \emph{Provenance-Aware Transformers}, an architecture that treats provenance as a first-class signal governing information flow within the model. 
As illustrated in Figure~\ref{fig:architecture}, three components realize this design: \emph{origin embeddings} encode the provenance of each token at the input layer; a learned \emph{origin attention bias} constrains cross-source influence directly within the attention mechanism; and a learnable \emph{origin scale}, paired with auxiliary geometric losses, preserves and regularizes provenance signal as it propagates through the forward pass. To adapt our architecture to pretrained models, we designed a two-stage supervised fine-tuning procedure: an \emph{origin fine-tuning} stage that teaches the model origin semantics across diverse tasks, followed by \emph{alignment fine-tuning} stage on a structural injection dataset that instills prompt-injection-resistant behavior.

Evaluation results on four IPI datasets indicate that Provenance-Aware Transformers remain
robust across both in-distribution and out-of-distribution settings, maintaining
near-zero attack success rate, and outperform a
broad range of existing defenses on out-of-distribution datasets.
We further evaluate the utility cost of our architecture and find that it preserves
general instruction-following capability comparable to the undefended base model,
indicating that robustness does not come at the expense of usefulness. More importantly, we show that our Provenance-Aware architecture remains robust against
white-box adaptive attacks explicitly optimized against the model, and that the same
suppression mechanism generalizes to other tasks such as jailbreak defense.

In summary, this paper makes the following contributions:
\begin{itemize}
  \item We propose \emph{Provenance-Aware Transformers}, a transformer architecture that establishes provenance as a first-class signal to enforce authoritative in provenance-aware LLM deployments.
  \item We design three purpose-built architectural components that enable the model to enforce provenance-based trust boundaries directly during generation.
  \item We develop a two-stage supervised fine-tuning pipeline that adapts our architecture to pretrained models.
  \item We construct two origin-labeled datasets for origin semantic learning and ring suppression behavior enforcement. 
  % \item We release our code and dataset to support future research on origin-aware and provenance-based defenses against prompt injection.
\end{itemize}

% =============================================================================
\section{Background \& Related Work}
\noindent \textbf{Decoder-Only LLMs.}
Modern LLMs predominantly adopt a decoder-only transformer architecture, in which each
token attends only to preceding tokens via causal self-attention, enabling
autoregressive generation one token at a time~\cite{radford2019gpt2,
vaswani2017attention}. This design has proven highly scalable, underlying models such as
GPT~\cite{brown2020gpt3}, LLaMA~\cite{touvron2023llama}, and
Mistral~\cite{jiang2023mistral}, and has become the dominant paradigm for
instruction-following and agentic LLM systems due to its simplicity and strong
generalization at scale. Decoder-only LLMs are susceptible to prompt injection because all input tokens are concatenated into a single causal sequence
and processed uniformly by this architecture\cite{zverev2025can}.

\noindent \textbf{Indirect Prompt Injection Attack.}
As LLMs are increasingly embedded into autonomous agents, retrieval pipelines, and enterprise applications\cite{claudecode,codex}, Indirect Prompt Injection has became a more realistic and concerning threat model where adversaries in which malicious instructions are not supplied
directly by the user but are instead smuggled into the model's context through external
content the application retrieves on the user's behalf. Formally, indirect prompt injection is defined as follows~\cite{greshake2023not,
yi2025benchmarking}: a user $u$ sends an instruction $I$
to an LLM-integrated application, which retrieves external content $C$ and combines it
with $I$ via a pre-defined prompt template $T$ to form a prompt $P = \mathit{Combine}(T,
C, f(I))$, where $f(I)$ denotes the instruction the application constructs from $I$ and
$\mathit{Combine}$ assembles the final prompt sent to the LLM to produce a response $R$.
% IPI arises when the external content $C$, rather than the user, is the source of
% adversarial instructions crafted to hijack the LLM's behavior away from the user's
% original intent. This threat model has evolved from simple instruction overrides to
% increasingly sophisticated exploits of model reasoning, and the broader body of attack
% techniques developed under the general PI remains directly applicable to the IPI
% setting, since $C$ can encode any of these techniques. 
Representative attacks include direct override and role-switching
patterns~\cite{Perez2022, Willison2023, Nakajima2022}, direct and indirect
injections embedded within retrieved content~\cite{liu2024promptinjection,
greshake2023not}, optimization-based attacks using adversarial suffixes or obfuscated
tokens~\cite{zou2023universal, zhang2025defense, Chen2024SecAlignDAA,
liu2025datasentinel, Hui2024}, and behavioral attacks exploiting model alignment through
persuasive or emotionally charged language~\cite{Shao2024, emotional_claude2024}.
% Early work showed LLMs could be coerced into ignoring system prompts through direct
% override or role-switching patterns~\cite{Perez2022, Willison2023, Nakajima2022}, later
% formalized into \emph{direct} overrides and \emph{indirect} injections embedded within
% retrieved content~\cite{liu2024promptinjection, Greshake2023}. Subsequent work has
% extended this threat model with optimization-based attacks that search for adversarial
% suffixes or obfuscated tokens~\cite{zou2023universal, zhang2025defense,
% Chen2024SecAlignDAA, liu2025datasentinel, Hui2024}, and behavioral attacks that exploit
% model alignment through persuasive or emotionally charged
% language~\cite{Shao2024, emotional_claude2024}, broadening PI from a narrow parsing
% exploit into a diverse class of semantic attacks on model reasoning.

\noindent \textbf{Indirect Prompt Injection Defense.}
Prior research on prompt injection defense could be generally categorized into
prompt-based, detection-based, and fine-tuning-based approaches. Although many of these
methods were proposed for prompt injection broadly rather than the indirect setting
specifically, they remain directly applicable to IPI, since indirect attacks differ only
in where the adversarial instruction originates ($C$), not in the mechanism by which it
must be neutralized. 
Prompt-based defenses operate at the input level without altering model weights, using
instructional reminders that direct the model to disregard embedded commands in external
content~\cite{Wallace2024InstructionHierarchy}, delimiter-based sandwich patterns~\cite{Willison2023,
learnprompting2023sandwich}, random
sequence framing that isolates untrusted spans with unpredictable
markers~\cite{Random_Sequnce}, and known-answer\cite{Nakajima2022} verification that probes for
injection by testing the model's response to content with a predetermined
answer~\cite{Nakajima2022} to enforce boundaries between instructions and untrusted
data, but remain vulnerable to
injected content that mimics legitimate instructions.
Detection-based approaches instead
treat prompt injection as a classification problem, ranging from low-latency heuristic
filters and classifiers like \textsc{PromptArmor}~\cite{shi2025promptarmorsimpleeffectiveprompt}
to generative guardrail models such as \textsc{Llama Guard}~\cite{inan2023llama}, semantic-intent detectors such as
\textsc{PromptSleuth}~\cite{wang2025promptsleuth}, game-theoretic detectors such as
\textsc{DataSentinel}~\cite{liu2025datasentinel}, and attention-based localization and
sanitization methods such as \textsc{PromptLocate}~\cite{jia2025promptlocate} and
\cite{zhong2025attention}, but operate downstream of generation with no architectural
guarantee against undetected injections. 
Fine-tuning-based approaches instead seek to internalize resistance directly into model
weights, through preference optimization as in \textsc{SecAlign}~\cite{Chen2024SecAlignDAA}, or
structured instruction-tuning as in \textsc{StruQ}~\cite{chen2025struq}, but rely entirely on the
base transformer's content-addressed attention to learn injection-resistant behavior from
examples, without any mechanism that structurally constrains how untrusted content can
influence generation.

Despite this progress, existing defenses remain fragmented and fundamentally reactive.
Prompt-based methods relying on structural patterns~\cite{chen2025struq, Willison2023} and
detection-based classifiers~\cite{inan2023llama} depend on rigid
lexical signatures, making them vulnerable to obfuscation and rephrasing.
Fine-tuning-based approaches~\cite{liu2025datasentinel, Chen2024SecAlignDAA,
chen2025struq} improve robustness but still operate at the content level, internalizing
resistance to specific attack patterns without addressing the architectural gap that makes
injection possible in the first place.
Across all existing approaches, the fundamental
problem remains: \emph{language models possess no mechanism to distinguish authoritative
instruction sources from non-authoritative data sources}~\cite{zverev2025can}, and
therefore have no structural basis for deciding what to obey.

Our work differs from these lines of defense in three important ways. \textbf{First},
unlike prompt-based methods such as delimiters, sandwiching~\cite{Willison2023, learnprompting2023sandwich}, we do not rely on the model to
infer from textual structure alone which spans are authoritative. Instead, provenance is
provided explicitly by the application and injected into the model as a first-class signal. 
\textbf{Second}, unlike detection-based approaches such as \textsc{DataSentinel}, and \textsc{PromptLocate}~\cite{liu2025datasentinel, jia2025promptlocate}, which operate as an external module outside the
protected model and offer no safety guarantee enforced on the model itself once an
injection evades detection, we build the defense into the protected model's own forward
pass, so that suppression of untrusted content is a property the model itself enforces
rather than a guarantee contingent on an external verification step succeeding. 
\textbf{Third}, unlike fine-tuning-based
approaches such as \textsc{StruQ} and \textsc{SecAlign}~\cite{chen2025struq, Chen2024SecAlignDAA}, our goal is not
merely to teach the model to resist known attack styles through training examples alone,
but to modify the forward pass itself so that source identity directly and structurally
constrains attention, independent of whether a given attack pattern resembles anything
seen during training.

% In this sense, Origin-Aware Transformers are best understood as a provenance-aware architectural defense. The method does not replace application-layer source control, nor does it solve semantic instruction detection in arbitrary text. Rather, it complements existing system defenses by giving the model an internal mechanism to act on provenance once that provenance is available. This shifts the core defense from content-level pattern matching toward source-level structural separation.

% =============================================================================
\section{Threat Model \& Problem Formulation}
\label{sec:threat-model}

We consider \emph{indirect prompt injection} in provenance-aware LLM applications, including
retrieval-augmented generation pipelines, tool-using agents, and document-grounded
assistants in which the application already mediates multiple input sources before
constructing the model context. In these systems, prompt injection arises when
adversarial text is embedded inside an external data source and is subsequently presented
to the model alongside legitimate instructions and user content\cite{greshake2023not, yi2025benchmarking}. In this section, we first
define the threat model between the adversary and defender, and then formulate the problem
of provenance-aware generation.

% \subsection{Adversary}
% We consider a standard IPI attack adversary\cite{greshake2023not, yi2025benchmarking} that manipulates external content sources, with the following specific goal, capabilities, and knowledge.

\noindent \textbf{Adversary's goal.} The adversary aims to redirect the LLM-integrated
application into executing attacker-chosen instructions, i.e., to exert
instruction-level control over assistant generation through content the adversary has
placed in an untrusted source.

\noindent \textbf{Adversary's capabilities.} We assume the attacker can control, partially
control, or poison an external content source that is later incorporated into the model
input, such as retrieved web pages, emails, documents, tickets, knowledge-base entries,
or tool-returned text. The attacker may use arbitrary phrasing, obfuscation, indirection,
or domain-specific formatting to disguise injected instructions. The attacker does
\emph{not} control the system prompt, the origin-assignment policy implemented by the
application, or the model weights at inference time.

\noindent \textbf{Adversary's knowledge.} 
We assume the adversary is aware that the defender may deploy an potential defense and can observe the LLM-integrated application's outputs in response to injected content. For our strongest, worst-case evaluation, we additionally
grant the adversary full white-box access to the target model's architecture, weights, and gradients, enabling gradient-based adaptive attacks optimized directly against our defense.
% We assume the adversary knows that the defender may
% adopt origin-aware defenses and may adapt phrasing accordingly. However, we do not
% assume the attack resembles any pattern seen during training, and the adversary has no
% knowledge of the specific origin labels assigned to its content beyond the fact that
% externally sourced data is treated as untrusted.

% \subsection{Defender}
% On the defender's side, we consider the following goal, capabilities, and knowledge.

\noindent \textbf{Defender's goal.} The defender's protected asset is \emph{instruction
  integrity}: the model should obey operator-issued policy and complete the
  user-authorized task without being redirected by untrusted external content. The
  defender aims to prevent tokens originating from untrusted sources from exerting
  instruction-level control over generation, while preserving the model's ability to use
  permitted sources as ordinary task context.

\noindent \textbf{Defender's capabilities.} 
We assume the application can assign an origin
  label to each token span before it is passed to the model. This does not require content-level judgment from the application-layer, the application already
  knows whether a token span originates from the system prompt, the assistant's prior
  output, direct user input, or an externally sourced document\cite{openai_prompt}. Also, we assume the defender can modify
  the model's architecture and fine-tune its weights, but cannot rely on the model to
  infer provenance from content.

\noindent \textbf{Defender's knowledge.} We assume the defender has white-box knowledge of the
target model and controls the origin-assignment policy at the application layer.
However, the defender does not know the exact wording, position, or obfuscation strategy
of any injected instruction in advance.

% \subsection{Problem Formulation}
\noindent \textbf{Problem Formulation.}
Given an input sequence assembled by the application, the tokenizer produces token
embeddings $\mathbf{F} = \mathcal{T}(X) = f_1, f_2, \dots, f_n$, where $f_i \in
\mathbb{R}^d$ and $\mathcal{T}$ is the tokenization function. In addition to the token
sequence, the application supplies an \emph{origin label} for every token, defined by a
mapping $\mathcal{O}: X \rightarrow \mathcal{R}^n$, where $\mathcal{R} = \{0, 1, 2, 3\}$
denotes the set of origin rings, as detailed in Section~\ref{sec:ring-hierarchy}. Concretely, $r_i = 0$ denotes operator-defined
instructions (Ring~0), $r_i = 1$ denotes the assistant's own prior tokens (Ring~1), $r_i =
2$ denotes direct user input (Ring~2), and $r_i = 3$ denotes structurally untrusted,
externally sourced content (Ring~3).

Rather than performing a binary judgment on the entire input $X$, as in existing
content-based defenses, we perform \emph{origin-conditioned generation}:
\begin{equation}
  \mathcal{M}_\Theta(\mathbf{F}, \mathbf{r}) = \mathcal{G}\Big(\big\{\mathcal{A}_\theta(f_i, r_i)\big\}_{i=1}^{n}\Big),
  \label{eq:origin-gen}
\end{equation}
where $\mathbf{r} = \mathcal{O}(X)$ is the origin-label sequence, $\mathcal{A}_\theta(\cdot)$
is our provenance-aware transformation that constrains how each token's representation may
influence subsequent generation as a function of its origin, and $\mathcal{G}(\cdot)$
aggregates these representations to produce the output sequence.

The security objective requires that tokens labeled as Ring~3 exert no instruction-level
influence on generation, regardless of their content. Let $I(f_i \rightarrow y)$ denote the
influence of token $f_i$ on output $y$ under $\mathcal{M}_\Theta$. We require:
\begin{equation}
  \forall f_i : r_i = 3, \quad
  I(f_i \rightarrow y) \not\models \texttt{instruction}(f_i),
  \label{eq:ring3-constraint}
\end{equation}
i.e., the model's output must not reflect instruction-level semantics carried by any
Ring~3 token, even when $I(f_i \rightarrow y) \neq 0$. Simultaneously, we require that origin-conditioning preserve
utility on permitted content:
\begin{equation}
  \forall f_i : r_i \in \{0, 1, 2\}, \quad
  \mathcal{M}_\Theta(\mathbf{F}, \mathbf{r}) \approx \mathcal{M}_{\Theta_0}(\mathbf{F}).
  \label{eq:utility-constraint}
\end{equation}
where $\mathcal{M}_{\Theta_0}$ denotes the base model's behavior absent origin
conditioning. Together, Equations~(2)--(4) capture the dual requirement of our defense:
suppressing instruction-level influence from Ring~3 while leaving task-relevant
information flow from Rings~0--2 intact.

\section{Provenance-Aware Transformer}

\subsection{Model Architecture}
The root vulnerability exploited by indirect prompt injection is that a language model
assigns no intrinsic meaning to the \emph{source} of a token~\cite{zverev2025can}. 
% Whether a token originates
% from an operator-defined system prompt, a user's input, or an adversary's payload hidden
% inside a retrieved document, the model processes it identically. 
This
\emph{instruction-data confusion} is a fundamental property of the standard transformer attention
mechanism, which is purely content-addressed: for tokens $t_i, t_j$ with hidden
representations $h_i, h_j$, the attention logit $a_{ij} = h_i^\top h_j / \sqrt{d}$ is a
function of content alone, with no term reflecting the provenance of either token.
% Surface-level defenses (delimiters, filters) can be bypassed by an adversary who crafts
% data that mimics the syntactic or semantic patterns of legitimate instructions.

In this paper we address this at the architectural level and propose \emph{Provenance-Aware Transformers}, which assigns every
input token $t_i$ a discrete \emph{Ring ID} $r_i \in \mathcal{R} = \{0,1,2,3\}$ that encodes its provenance
before the first transformer layer, and enforce this provenance structurally as the token
propagates through the model. Realizing this provenance-aware defense, however, introduces
three challenges that mirror those motivating this work: the model must be taught what an
origin label actually means; that meaning must survive the network's normalization and
content-addressed attention long enough to matter; and the resulting suppression must not
degrade behavior on benign, permitted content. We correspondingly propose three origin components added to the decoder-only transformer architecture\cite{vaswani2017attention}, as illustrated in Figure~\ref{fig:architecture}.

\textbf{First}, assigning a ring ID to a token does not by itself teach the model what that
label implies about trust. To mitigate this, we designed an \emph{origin embedding} table
$\mathbf{E}_{\text{org}} \in \mathbb{R}^{4 \times d}$ indexed by ring ID and added to each
token's standard embedding before it enters the transformer stack, as illustrated in
Figure~\ref{fig:architecture}. The application layer tags every token in the input sequence
with its ring ID (Ring Assignment), which is looked up in the origin embedding table and
merged into the token's representation (Origin Embedding). Because this embedding is a
learned parameter rather than a static tag, the two-stage training procedure of
Section~\ref{sec:training} can shape it to carry real, source-specific semantics: Stage~1
origin fine-tuning teaches the model what each ring implies about trust before any
suppression behavior is trained, giving the label the content that assigning it alone
could not provide.

\textbf{Second}, even a correctly learned origin signal must persist through the forward
pass to have any effect on generation, and standard transformer operations provide no
guarantee of this (especially considering the dense layers nowadays transformer models
have). To mitigate this, we designed two complementary mechanisms. An \emph{origin scale}
$\alpha \in \mathbb{R}^+$, shown in Figure~\ref{fig:architecture}, multiplies the origin
embedding before it is summed with the token embedding; because $\alpha$ is itself
learnable, the model can grow it to counteract the immediately downstream normalization
layer, which would otherwise dilute the origin signal's magnitude as token embeddings grow
during fine-tuning. An \emph{origin attention bias} then gives source identity a direct
channel inside the attention computation itself: as shown in Figure~\ref{fig:architecture},
a learned bias term $b(r_i, r_j)$, indexed by the ring IDs of the query and key tokens, is
added to the raw attention logits, so origin constrains the model's behavior at every layer
even though content-addressed attention alone provides no native mechanism to do so.

\begin{figure*}[!t]
\centering
\includegraphics[width=0.8\textwidth]{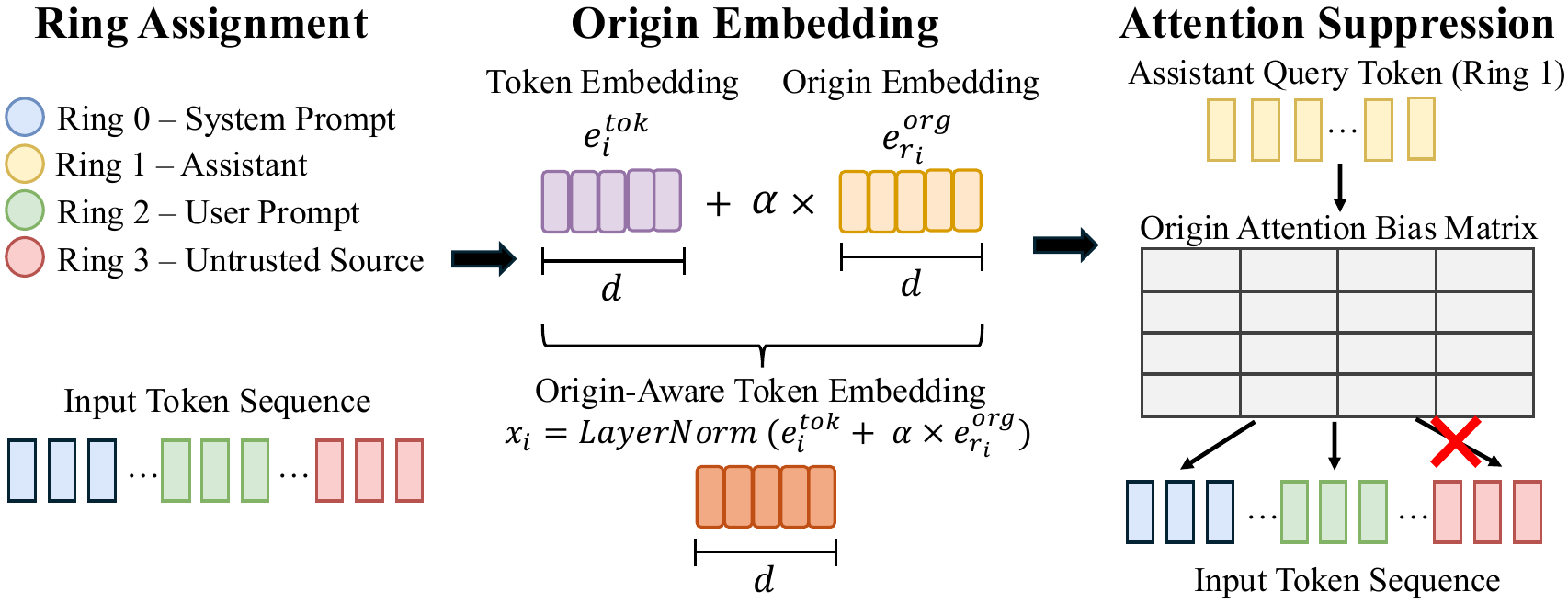}
\caption{Overview of Provenance-Aware Transformer. Each input token is tagged with a ring ID
encoding its provenance. The ring ID contributes an additive origin embedding to the token
representation. Within each attention layer, a learned bias matrix modulates attention
logits based on the ring IDs of the query and key tokens.}
\label{fig:componants}
\end{figure*}

\textbf{Third}, the suppression mechanism must not degrade model behavior on benign,
permitted content. The origin attention bias handles part of this by being selective: only
the bias term $b(1, 3)$, governing attention from the assistant's own queries (Ring~1) to
untrusted keys (Ring~3), is strongly suppressive, while $b(1, 2)$ and other permitted pairs
are left near zero. As Figure~\ref{fig:componants} shows, this means attention to permitted
Ring~2 content is left unaffected, so ordinary task-relevant information keeps flowing
normally.
Furthermore, the added components perturbs the
transformer's internal latent space by alter token representations and attention patterns that the
model's original weights were optimized around, so the model must relearn how to perform
ordinary tasks in the presence of these new signals to preserve utility. We therefore designed a two-stage training pipeline, illustrated in
Figure~\ref{fig:pipeline}: an origin fine-tuning stage first exposes the model to a broad,
diverse instruction-following corpus so that competence with the added architectural
components is re-established before any suppression objective is introduced, and only
then does an alignment fine-tuning stage teach Ring~3 suppression on a narrow structural
injection dataset. Evaluation results in Section~\ref{sec:rq3-ref} show that our model achieves around 50\% LC win
rates against the base pretrained model on AlpacaEval 2.0\cite{alpaca_eval}, indicating equivalent utility performance to base pretrained model\cite{Chen2024SecAlignDAA,zhong2025attention}.
% To further prevent downgrading model utility, we designed a
% two-stage training pipeline, illustrated in Figure~\ref{fig:pipeline}: an origin
% fine-tuning stage first exposes the model to a broad, diverse instruction-following corpus
% so that general capability is established before any suppression objective is introduced,
% and only then does an alignment fine-tuning stage teach Ring~3 suppression on a narrow
% structural injection dataset. Because the model's general competence is anchored before the
% narrow stage begins, the defense does not guarantee security at the cost of usefulness.

\begin{table}[!t]
\renewcommand{\arraystretch}{1.2}
\caption{The Four-Ring Trust Hierarchy}
\label{tab:rings}
\centering
\footnotesize
\begin{tabular}{clll}
\toprule
\textbf{ID} & \textbf{Label} & \textbf{Source} & \textbf{Role} \\
\midrule
0 & \texttt{ORG\_SYSTEM} & Operator    & \makecell[c]{Absolute authority that \\ defines policy} \\
1 & \texttt{ORG\_SELF}   & Model       & \makecell[c]{Assistant-generated tokens} \\
2 &  \texttt{ORG\_USER}   & User / Trusted Data & \makecell[c]{User intent and other \\ permitted/trusted content}\\
3 & \texttt{ORG\_Data} & Untrusted Source   & \makecell[c]{Untrusted data source \\ that is potentially injected} \\
\bottomrule
\end{tabular}
\end{table}

\subsection{Ring Hierarchy}
\label{sec:ring-hierarchy}
Following standard prompt structure~\cite{openai_prompt}, we define four rings organized
into a strict trust hierarchy, summarized in Table~\ref{tab:rings}. Ring~0
(\texttt{ORG\_SYSTEM}) represents the operator and carries absolute authority: it defines
the policy the model must unconditionally obey. Ring~1 (\texttt{ORG\_SELF}) is reserved for
the model's own generated tokens, allowing the model to reason over and build upon its
prior output without conflating it with externally supplied instructions. Ring~2
(\texttt{ORG\_USER}) encompasses direct user intent along with other content the
deployment designates as trusted or permitted, which the model processes as ordinary task
context. Ring~3 (\texttt{ORG\_Data}) is reserved for untrusted, potentially adversarial
data sources; content assigned to this ring is treated as structurally non-authoritative
and is suppressed unconditionally, regardless of its surface content.

We would like to point out that in practice, deployments with more complex trust requirements can
define finer-grained hierarchies that extend beyond these four rings. For instance,
distinguishing between multiple classes of trusted external sources with different
privilege levels, or separating tool-returned content from retrieved documents even when both are structurally untrusted. The architecture itself imposes no constraint on the number of rings; it is the application, not the model, that determines how many
provenance classes are meaningful for a given deployment.

\subsection{Ring Assignment}
\label{sec:assignment}

Ring IDs are assigned at the application layer before any model computation. The
assignment maps each token span to a ring based on its structural role in the input:
\begin{equation}
r_i =
\begin{cases}
0 & t_i \in \text{system prompt} \\
1 & t_i \in \text{previously generated assistant tokens} \\
2 & t_i \in \text{user input or other permitted content} \\
3 & t_i \in \text{structurally untrusted external content}
\end{cases}
\label{eq:assign}
\end{equation}
Assignments to Ring~0 and Ring~2 do not require an explicit detection mechanism, since the application framework already has access to the provenance of these inputs\cite{openai_prompt}. System instructions are supplied directly by the framework, whereas user inputs are received through the user-facing interface. In the deployment setting considered in this work, Ring~3 is likewise determined by provenance policy and does not require inspection to the semantic content of the input. A straightforward policy is to assign all externally sourced content that is not structurally trusted to Ring~3, thereby treating such content as non-authoritative by default. More fine-grained, span-level provenance annotations could be incorporated as an engineering optimization, but they are not required by our core design. The key observation is that, once provenance information is explicitly represented, the model can enforce the corresponding trust boundaries structurally, without relying on content-based detection of potentially adversarial instructions.

\subsection{Origin Components Design}
\label{sec:arch}

Figure~\ref{fig:architecture} illustrated the proposed Provenance-Aware Transformer architecture. As Figure~\ref{fig:architecture} shows, we introduce three trainable components on top of the decoder-only transformer architecture: (1) a origin embedding that embeds the origin ring label; (2) a learnable gate for gated origin integration with token embedding; (3) a origin-aware attention bias matrix to enforce suppression based on origin embedding. In this section, we describe in more details about the design for each origin component, as illustrated in Figure~\ref{fig:componants}.

\noindent \textbf{Origin Embedding.}
Let $\mathbf{E}_{\text{tok}} \in \mathbb{R}^{|V| \times d}$ be the standard token
embedding table for vocabulary $V$ and model dimension $d$. We introduce an additional
embedding table $\mathbf{E}_{\text{org}} \in \mathbb{R}^{4 \times d}$ indexed by ring ID.
As Figure~\ref{fig:componants} shows, for token $t_i$ at position $i$ with ring ID $r_i$, the input representation is:
\begin{equation}
  \mathbf{x}_i = \mathbf{E}_{\text{tok}}[t_i] + \alpha \cdot \mathbf{E}_{\text{org}}[r_i]
  \label{eq:embed}
\end{equation}
where $\alpha \in \mathbb{R}^+$ is a learnable scalar gate initialized to $\alpha_0 = 0.1$. By introducing this origin embedding, we incorporate origin information directly into the
model's input representation, ensuring that provenance is available to every downstream
computation from the very first layer.

\noindent \textbf{Origin Integration Gate.}
The RMSNorm layer applied immediately after Eq.~\eqref{eq:embed} normalizes the combined
representation, and during subsequent fine-tuning on large corpora the token embedding
$\mathbf{E}_{\text{tok}}$ can grow in magnitude, progressively diluting the relative
contribution of the origin signal to the point where only its direction (not its
magnitude) survives normalization. We therefore introduce a gated origin
integration gate $\alpha$ that assists in preserving the origin ring label for consequent calculations,
helping the origin information to survive through dense attention blocks. 
% Making $\alpha$ a trainable parameter allows the model
% to continuously recalibrate the balance between token semantics and origin information as
% training progresses. 
While sufficient training could in principle learn this balance
implicitly through the embeddings alone, an explicit scalar gate is a far easier
optimization target, yielding more stable and faster convergence than requiring
high-dimensional embedding weights to compensate for a scale mismatch.

% \noindent \textbf{Origin Integration Gate.}
% The RMSNorm layer applied
% immediately after Eq.~\eqref{eq:embed} normalizes the combined representation, and during
% subsequent fine-tuning on large corpora the token embedding $\mathbf{E}_{\text{tok}}$ can
% grow in magnitude, progressively diluting the relative contribution of the origin signal
% to the point where only its direction (not its magnitude) survives normalization.
% Making $\alpha$ a trainable parameter allows the model to continuously recalibrate the
% balance between token semantics and origin information as training progresses. This gated origin integration assists in preserving the origin ring label for consequent calculations, helping the origin information to survive through dense attention blocks.

\begin{figure*}[!t]
\centering
\includegraphics[width=0.8\textwidth]{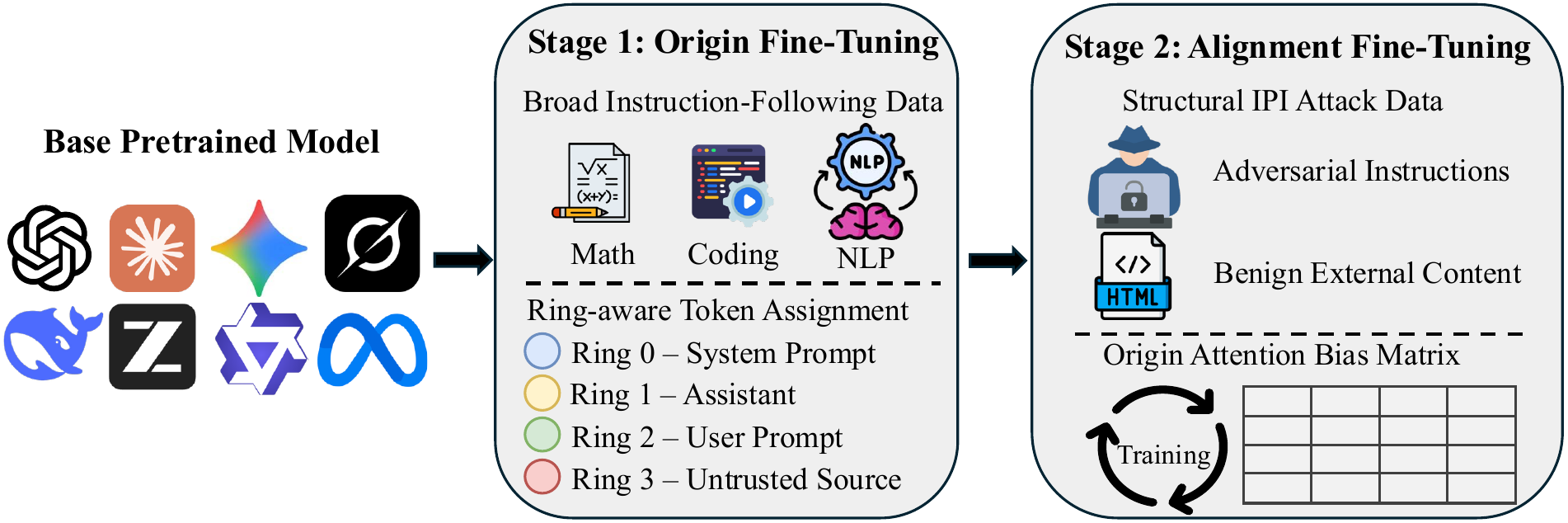}
\caption{Two-stage supervised fine-tuning methodology for adapting Provenance-Aware
Transformers to pretrained checkpoints. Stage~1 teaches provenance semantics across broad
instruction-following data, and Stage~2 aligns behavior on the structural injection
dataset under ring constraints.}
\label{fig:pipeline}
\end{figure*}

\noindent \textbf{Origin Attention Bias.}
While the previous components added and preserves the origin ring labels in the model, the core defense mechanism is a learned bias matrix $B \in \mathbb{R}^{4 \times 4}$
applied to attention logits, as illustrated in Figure~\ref{fig:componants}. In standard scaled dot-product attention, the logit between
query position $i$ and key position $j$ is $\mathbf{q}_i \cdot \mathbf{k}_j / \sqrt{d_k}$.
Because $B$ is indexed by ring rather than by position, applying it across a sequence of
length $T$ requires expanding it according to the ring-ID sequence: for every query-key
position pair $(i,j)$, the model gathers the entry $B_{r_i, r_j}$ corresponding to the ring
IDs of the query and key tokens, producing a $T \times T$ bias tensor that is added
elementwise to the raw attention scores before softmax. The resulting per-pair logit is:
\begin{equation}
  a_{ij} = \frac{\mathbf{q}_i \cdot \mathbf{k}_j}{\sqrt{d_k}} + B_{r_i,\, r_j}
  \label{eq:attn}
\end{equation}
where $B_{r_i, r_j}$ denotes the entry of $B$ obtained by this gather operation rather than
by matrix multiplication. The matrix $B$ is shared across all attention layers and heads
and is trained end-to-end.

Rather than fixing specific numeric values in the paper, which are empirically-tuned
hyperparameters and not part of the architectural contribution, we describe $B$'s
initialization at the alignment training stage by its qualitative policy and optimize this bias matrix through the training process:
\begin{equation}
B_{\text{init}} =
\begin{blockarray}{ccccc}
  & \scriptstyle r_j{=}0 & \scriptstyle r_j{=}1 & \scriptstyle r_j{=}2 & \scriptstyle r_j{=}3 \\
\begin{block}{c(cccc)}
\scriptstyle r_i{=}0 \; & 0 & 0 & 0 & 0 \\
\scriptstyle r_i{=}1 \; & 0 & 0 & 0 & -\delta \\
\scriptstyle r_i{=}2 \; & 0 & 0 & 0 & -\delta \\
\scriptstyle r_i{=}3 \; & 0 & 0 & 0 & 0 \\
\end{block}
\end{blockarray}
\label{eq:bias_init}
\end{equation}
where $\delta \gg 0$ is a fixed constant. Only Ring~1 (assistant) and Ring~2 (user)
queries are suppressed when attending to Ring~3 (untrusted) keys; every other query-key
relationship is left unrestricted. This covers both directions through which untrusted
content could otherwise exert influence: directly, when the assistant's own queries attend
to untrusted keys during generation, and indirectly, when permitted Ring~2 content attends
to untrusted keys and could carry that influence forward into representations the assistant
later builds upon. 
Ring~0 queries require no explicit suppression term, since the system prompt is always placed before any untrusted content in the token sequence and causal masking alone already prevents system-position queries from attending to untrusted keys.
Ring~3's own queries are left unrestricted, since constraining what
untrusted content itself attends to has no bearing on the final output unless a later,
unsuppressed query attends back to it. This attention bias matrix creates a physical-level separation between untrusted sources and other sources, providing a architectural level defense guarantee. To correctly optimize the matrix, we further proposed two auxiliary loss functions, as illustrated below.

\subsection{Training Pipeline for Pretrained Models}
\label{sec:training}

To adapt our proposed model architecture to existing released pretrained model checkpoints, we
propose a practical two stage supervised-fine-tuning pipeline, as illustrated in
Figure~\ref{fig:pipeline}. We further propose two customized auxiliary loss functions to enforce our suppression policy during backpropagation, as described in Equation~\ref{eq:l_orth} and Equation~\ref{eq:l_auth}.

\smallskip\noindent\textbf{Stage 1 --- Origin Fine-Tuning.}
The pretrained model is first fine-tuned on a broad instruction-following corpus with
proper ring assignments. This stage is designed to teach the new architectural components
the semantics of source provenance before the model is asked to resist prompt injection on
a narrow alignment dataset. Two auxiliary losses regularize the origin embeddings, where
$\mathbf{e}_r = \mathbf{E}_{\text{org}}[r]$ denotes the origin embedding for ring $r$. An
orthogonality loss penalizes cosine similarity between the Ring~0 and Ring~2 embeddings,
\begin{equation}
  \mathcal{L}_{\text{orth}} = \left( \frac{\mathbf{e}_0 \cdot \mathbf{e}_2}{\|\mathbf{e}_0\| \, \|\mathbf{e}_2\|} \right)^{2},
  \label{eq:l_orth}
\end{equation}
encouraging the system-prompt and user embeddings to occupy geometrically distinct
directions. An authority loss additionally encourages the system-prompt embedding to
dominate in magnitude,
\begin{equation}
  \mathcal{L}_{\text{auth}} = \text{ReLU}\big(\|\mathbf{e}_2\| - \|\mathbf{e}_0\| + m\big),
  \label{eq:l_auth}
\end{equation}
where $m \geq 0$ is a margin. Together, these losses encourage the model to form
geometrically distinct representations for the principal ring classes before the
task-specific alignment stage begins. Conceptually, Stage~1 teaches the pretrained model
how to represent and preserve provenance, not just how to imitate task outputs.

\smallskip\noindent\textbf{Stage 2 --- Alignment Fine-Tuning.}
Following origin fine-tuning, the model is fine-tuned on a purpose-built structural injection dataset that provides explicit supervision for the suppression behavior required by the architecture. 
% At the outset of this stage, the attention bias matrix $B$ is reset to $B_{\text{init}}$
%   (Eq.~\eqref{eq:bias_init}), overriding drifts accumulated during Stage~1 and ensuring that alignment fine-tuning begins from a configuration already consistent with Ring~3-adversarial semantics rather than from the task-following bias induced by origin fine-tuning. 
  This stage instills the behavioral
  policy associated with the ring hierarchy: Ring~0 instructions are followed unconditionally, Ring~2 content is processed as ordinary task input, and Ring~3 content is suppressed regardless of its surface form.

  Teaching this policy requires training examples in which provenance rather than content determines the correct behavior, a property that generic instruction-tuning corpora and existing prompt-injection datasets do not provide, since neither annotates which spans originate from which source. We
  therefore construct a purpose-built dataset by extending the witness-based methodology of the SEP benchmark~\cite{zverev2025can}, which embeds a verifiable secret instruction inside otherwise-benign data to test whether a model executes content it should only process, with explicit ring provenance labels
  consistent with the ring hierarchy in Table~\ref{tab:rings}. Detailed implementation of the dataset is discussed in Section~\ref{sec:experiment_setup}.

% =============================================================================
\section{Evaluation}
% We evaluate Origin-Aware Transformers as a \emph{provenance-aware defense} against
% indirect prompt injection. Our evaluation is designed to answer six questions: how
% Origin-Aware compares to existing defenses on a combined benchmark, whether origin-aware
% conditioning improves robustness to unseen prompt injection styles, whether it preserves
% utility on benign tasks, whether the gain is structural rather than merely
% in-distribution memorization, how it holds up against an adaptive adversary, and whether
% Ring~2 suppression generalizes beyond prompt injection to a different threat class
% (jailbreaking).

We evaluate our Provenance-Aware Transformer seeking answers to the following research questions:

\smallskip\noindent\textbf{RQ1 (Effectiveness): What is the effectiveness in defending against indirect prompt injection of Provenance-Aware Transformer compared to existing work?}

\smallskip\noindent\textbf{RQ2 (Generalizability): Does Provenance-Aware Transformer generalize to different model and out-of-distribution attacks?}

\smallskip\noindent\textbf{RQ3 (Utility): Does the defense preserve benign-task utility?}

\smallskip\noindent\textbf{RQ4 (Adaptive Adversarial): How does Provenance-Aware Transformer perform under adaptive indirect prompt injection attacks?}

\smallskip\noindent\textbf{RQ5 (Ablation Study): How much does each component contribute in the Provenance-Aware Transformer architecture?}

\smallskip\noindent\textbf{RQ6 (Applications Scenarios): Does our model architecture generalize to other threat class targeting safety alignment other than indirect prompt injection?}

\subsection{Experimental Setup}
\label{sec:experiment_setup}

\smallskip\noindent\textbf{Implementation.}
Our provenance-aware architecture is implemented as a lightweight wrapper around pretrained HuggingFace language models, where we extend a pretrained checkpoint with the three learned
  components described in Section~\ref{sec:arch}. The origin attention bias is applied by gathering
  ring-indexed bias values into a full attention-score-shaped tensor and injecting it as an additive term into the causal attention mask, allowing it to be computed via the standard scaled-dot-product-attention kernel without a custom attention implementation. We then utilize the data-pipeline and training-loop infrastructure from nanochat\cite{nanochat}, a from-scratch GPT-style transformer codebase, to run the two-stage training procedure described in
  Section~\ref{sec:training} that adapts the wrapped checkpoint into an provenance-aware, injection-resistant model.
  All training is performed with DeepSpeed
  ZeRO-2 across 8~NVIDIA A100 GPUs (80GB each), which shards optimizer state and gradients across devices to accommodate the 7--8B-parameter models in our evaluation. Stage~1 (origin fine-tuning) uses a per-GPU batch size of 2 with 8 gradient-accumulation steps, a maximum
  sequence length of 1024 tokens, learning rate $1\times10^{-5}$, and gradient clipping at norm~1.0, for 2 epochs. Stage~2
  (alignment fine-tuning) uses the same per-GPU batch size and accumulation schedule with a maximum sequence length of 2048 tokens, learning rate $2\times10^{-5}$, for 3 epochs.

\smallskip\noindent\textbf{Base Model.}
We evaluate our architecture on four openly released decoder-only models: \textit{SmolLM2-360M-Instruct}\cite{allal2025smollm2smolgoesbig}, \textit{LLaMA-3-8B-Instruct}\cite{llama3modelcard}, \textit{Qwen2.5-7B-Instruct}\cite{qwen2.5}, \textit{Mistral-7B-Instruct-v0.3}\cite{mistral-7b}. We consider \textit{LLaMA-3-8B-Instruct} as our primary evaluation model, several baseline defenses we compare against (Section~\ref{sec:rq1}) were themselves designed or officially released for the LLaMA family, which maximizes comparability. \textit{SmolLM2-360M-Instruct} is included to enable fast iteration during development and test whether the architecture's structural generalization property holds even at a scale far below typical production LLMs.

\smallskip\noindent\textbf{Baselines.}
For RQ1, we compare against ten defenses spanning three categories, in addition to an undefended baseline.
  \emph{Prompt-based defenses} intervene only at the input level, without modifying model weights: Instructional prompting~\cite{Wallace2024InstructionHierarchy}, the Sandwich defense~\cite{learnprompting2023sandwich}, Random Sequence disclosure~\cite{Random_Sequnce}, Delimiter-based
  isolation~\cite{Willison2023}, and Known-Answer detection~\cite{Nakajima2022}.
  \emph{Detection-based defenses} classify inputs as injected or benign prior to generation: PromptSleuth~\cite{wang2025promptsleuth}, DataSentinel~\cite{liu2025datasentinel}, PromptGuard~\cite{promptguard}, PromptLocate~\cite{jia2025promptlocate}, and Rennervate~\cite{zhong2025attention}.
  \emph{Fine-tuning-based defenses} modify model weights: SecAlign~\cite{Chen2024SecAlignDAA}, which applies preference optimization via officially released LoRA adapters, and our own Architecture-Free Baseline (fine-tuning only, no ring components), which isolates the effect of alignment
  fine-tuning alone from the proposed architecture.

\smallskip\noindent\textbf{Training Datasets.}
\emph{Stage 1 --- Origin Fine-Tuning Data.}
  Stage~1 uses a 105K-example corpus assembled from four public sources, each contributing a distinct task format: UltraChat-200K (80K examples), a large multi-turn conversational instruction-following dataset providing broad, general-purpose linguistic coverage; MMLU auxiliary-train (12K), multiple-choice
  academic-knowledge questions spanning many domains, providing structured, classification-style reasoning distinct from open-ended dialogue; GSM8K (8K), grade-school math word problems, providing multi-step quantitative reasoning; and CodeAlpaca-20K (5K), code-instruction-following pairs, providing a
  structurally distinct symbolic domain. We selected these four sources jointly to maximize task-format diversity to satisfy Stage~1's role (Section~\ref{sec:arch}) to anchor general capability broadly before the
  narrow Stage~2 objective is introduced.
  
\emph{Stage 2 --- Structural IPI Dataset.}
  We construct the Stage~2 dataset by extending the witness-based methodology of the SEP benchmark~\cite{zverev2025can} with explicit ring provenance labels, following four steps.
  
  \emph{Step 1 --- Context generation.} For each of SEP's task subtasks (defined by a name, description, and system prompt), we use Claude to generate new short data contexts consistent with that subtask, deduplicated against existing SEP content, expanding coverage beyond the benchmark's original fixed
  set.
  
  \emph{Step 2 --- Ground truth.} For each generated context, we separately prompt Claude with only the legitimate task instruction and the context to produce a one-sentence reference response; this becomes the supervised target for that example regardless of whether an injection is later added.
  
  \emph{Step 3 --- Probe selection.} We sample a probe--witness pair from a fixed bank: each probe is an adversarial instruction (e.g., an override directive) paired with a deterministic witness string that would only appear in the model's output if the probe were obeyed instead of the real task, providing
  a verifiable, rule-based signal of compromise.
  
  \emph{Step 4 --- Injection construction.} We build on PromptSleuth-Bench~\cite{wang2025promptsleuth} and included 9 attack patterns to construct paired examples per context: an \emph{injection} example,
where a probe is inserted at a randomized position within the Ring~2 context and tagged
Ring~3, supervised with the Step~2 ground truth (i.e., the injection must be ignored); and
a matched \emph{benign} example with identical context and ground truth but no probe,
ensuring identical supervision regardless of whether adversarial content is present. We
supplement this with general-purpose examples from Alpaca-Cleaned~\cite{alpaca}, converted
to clean Ring~0/1/2 conversations with no Ring~3 content, anchoring benign behavior with
both SEP-derived and general instruction-following data. This yields 60,200 training
examples (24,079 injection, 36,121 benign) and a held-out IID test set of 4,375 examples.
  % \emph{Step 4 --- Injection construction.} We build upon PromptSleuth-Bench\cite{wang2025promptsleuth}, which includes 9 distinct prompt injection attack patterns and build two paired examples per context: an \emph{injection} example, in which the probe is appended to the context (with randomized insertion position) inside the user turn and tagged Ring~3, while the context itself remains Ring~2, supervised with the Step~2
  % ground truth (i.e., the injection is present but must be ignored); and a matched \emph{benign} example with the same context and ground truth but no probe present, ensuring the model is supervised identically whether or not adversarial content happens to appear. This benign supervision is further supplemented with general-purpose examples sampled from Alpaca-Cleaned\cite{alpaca},
  % converted to clean Ring~0/Ring~2/Ring~1 conversations with no Ring~3 content, so that the model's non-adversarial behavior is anchored by both SEP-derived and general-purpose instruction-following data. This process yields a training set containing 60,200 examples (24,079 injection, 36,121 benign) and a
  % held-out test set of 4,375 examples used for IID evaluation.

\smallskip\noindent\textbf{Evaluation Datasets.} To evaluate if our proposed defense generalizes structurally rather than merely
  fitting the training distribution, we evaluate on one Independent and Identically Distributed (IID) set and three Out-of-Distribution (OOD) sets, summarized in Table~\ref{tab:eval_datasets}. The IID set (SEP test~\cite{zverev2025can}) is the held-out split from the same benchmark used to construct our Stage~2 training data, and measures
  whether a model fits the alignment distribution it was trained on. The three OOD sets PI-Attack~\cite{pi-attack}, DataSentinel~\cite{liu2025datasentinel}, and Math-Tutor~\cite{wang2025promptsleuth} are drawn from different task domains and injection styles than SEP, and were never seen during training.

\smallskip\noindent\textbf{Metrics.} We mainly consider two metrics to measure the effectiveness of a defense to IPI attack and the utility of a fine-tuned model.

\smallskip\noindent\textbf{Attack Success Rate (ASR).} For a set of injection examples $D_{\text{inject}}$, each paired with the label $y_{\text{inject}}$ implied by the injected instruction, ASR is the fraction of examples for which the model's output matches or reflects that label rather than the
  correct task label:
  $$\text{ASR} = \frac{|{x \in D_{\text{inject}} : \text{label}(f(x)) = y_{\text{inject}}(x)}|}{|D_{\text{inject}}|}.$$

  \smallskip\noindent\textbf{Win Rate.} We evaluate general-purpose utility using AlpacaEval~2.0~\cite{alpaca_eval} (805 open-ended instruction-following prompts). For each prompt $x$, we generate a response from the candidate model $f(x)$ and from the base pretrained model $f_{\text{base}}(x)$, then present both to an LLM judge (Claude
  Sonnet~5) in randomized order (to cancel position bias) and record which response is preferred; both responses are truncated to a matched character budget before judging to prevent the judge from simply preferring the longer response. The raw win rate over the prompt set $D$ is
  $$\text{Win Rate} = \frac{1}{|D|}\sum_{x \in D} \mathbb{1}\left[\,j(f(x), f_{\text{base}}(x)) = \text{candidate}\,\right],$$
  where $j(\cdot,\cdot) \in \{\text{candidate}, \text{base}\}$ denotes the judge's preference between the two responses for a given prompt. We report the \emph{length-controlled} (LC) win
  rate~\cite{alpaca_eval_lc}, which corrects for residual verbosity bias by fitting a logistic model of the judge's preference as a function of a length-difference term, $\text{logit } P(\text{candidate} \succ \text{reference}) = \alpha + \beta \cdot \tanh(\Delta_{\text{len}} /
  \sigma_{\Delta_{\text{len}}})$, then reporting $\text{LC win rate} = \sigma(\alpha)$ with the length-sensitivity term $\beta$ zeroed out; a value of 50\% indicates parity with the undefended base model.

\begin{table}[!t]
\renewcommand{\arraystretch}{1.2}
\caption{Evaluation datasets used in the paper.}
\label{tab:eval_datasets}
\centering
\footnotesize
% \begin{tabular}{llp{3.4cm}r}
\begin{tabular}{cccc}
\toprule
\textbf{Dataset} & \textbf{Setting} & \textbf{Task Type} & \textbf{Size} \\
\midrule
SEP\cite{zverev2025can}                & IID & \makecell[c]{Witness-based injection \\ (multi-domain)}          & 4{,}375 \\
PI-Attack\cite{pi-attack}              & OOD & \makecell[c]{Classification \\(sentiment/spam/NLI/hate)}        & 8{,}256 \\
DataSentinel\cite{liu2025datasentinel} & OOD & \makecell[c]{Classification \\(sentiment/similarity/NLI)}       &    178 \\
Math-Tutor\cite{wang2025promptsleuth}  & OOD & \makecell[c]{Free-form math \\ problem solving}                  & 3{,}800 \\
\bottomrule
\end{tabular}
\end{table}

\begin{table*}[!t]
\renewcommand{\arraystretch}{1.2}
\caption{Comparison of Provenance-Aware Transformer against existing defenses across all
evaluation datasets (LLaMA-3-8B-Instruct). \textbf{ASR\,\%}: attack success rate on
injection examples (lower is better); for detection-based methods, ASR equals the
false-negative rate (fraction of injections not flagged). \textbf{Bold} marks the
best value per metric and dataset once all results are available.}
\label{tab:comparison}
\centering
\small
\begin{tabular}{ll cccc}
\toprule
\textbf{Category} & \textbf{Method} &
  \textbf{SEP (IID)} & \textbf{PI-Attack} & \textbf{DataSentinel (OOD)} & \textbf{Math-Tutor (OOD)} \\
 & & \textbf{ASR\%} & \textbf{ASR\%} & \textbf{ASR\%} & \textbf{ASR\%} \\
\midrule
None & No Defense
  & 32.5 & 21.3 & 24.1 & 17.2 \\
\midrule
\multirow{5}{*}{Prompt-based}
  & Instructional\cite{Wallace2024InstructionHierarchy}
  & 30.9 & 16.1 & 20.7 & 19.9 \\
  & Sandwich\cite{learnprompting2023sandwich}
  & 25.5 & 4.6 & 13.8 & 13.2 \\
  & Random Sequence\cite{Random_Sequnce}
  & 15.8 & 2.1 & 12.1 & 26.6 \\
  & Delimiters\cite{Willison2023}
  & 11.7 & 9.0 & 22.4 & 4.1 \\
  & Known-Answer\cite{Nakajima2022}
  & 10.9 & 5.4 & 3.4 & 10.1 \\
\midrule
\multirow{5}{*}{Detection-based}
  & PromptSleuth\cite{wang2025promptsleuth}
  & 9.6 & 23.6 & 5.2 & 4.3 \\
  & DataSentinel\cite{liu2025datasentinel}
  & 0.5 & 0.0 & 1.7 & 2.3 \\
  & PromptGuard
  & 28.0 & 53.8 & 50.0 & 35.0 \\
  & PromptLocate\cite{jia2025promptlocate}
  & 0.0 & 2.0 & 0.0 & 8.0 \\
  & Rennervate\cite{zhong2025attention}

  & 5.5 & 7.5 & 11.6 & 0.9 \\
\midrule
\multirow{3}{*}{Fine-tuning}
  & SecAlign\cite{Chen2024SecAlignDAA}
  & 2.0 & 0.0 & 6.9 & 0.0 \\
  & Baseline SFT
  & 0.8 & 2.3 & 8.6 & 0.3 \\
  & \textbf{Provenance-Aware (Ours)}
  & \textbf{0.6} & \textbf{0.0} & \textbf{0.0} & \textbf{0.0} \\
\bottomrule
\end{tabular}
\end{table*}

\subsection{RQ1: Comparison to Existing Defenses}
\label{sec:rq1}

Table~\ref{tab:comparison} shows that no existing defense category achieves consistently
low ASR across all evaluation settings. Prompt-based methods provide only modest IID gains
(10.9\%--30.9\% versus 32.5\% with no defense) and largely fail to transfer OOD: Random
Sequence, the strongest prompt-based method, still reaches 12.1\% and 26.6\% ASR on
DataSentinel and Math-Tutor, respectively. These reformulations shift word order or
add markers but leave the injected instruction semantically intact, giving the model no
principled basis for ignoring it under novel phrasing. Detection-based defenses show wide
variance: DataSentinel and PromptLocate drive OOD ASR near zero, while PromptGuard remains
high across all four datasets (28.0\%--53.8\%) and PromptSleuth reaches 9.6\% on SEP.
Among fine-tuning methods, SecAlign achieves competitive OOD ASR on PI-Attack and
Math-Tutor (0.0\% each) but rises to 6.9\% on DataSentinel. Provenace-Aware achieves 0.00\% ASR across all three OOD datasets, matching or exceeding the
strongest result in every category and establishing state-of-the-art robustness against
indirect prompt injection.

\begin{table*}[!t]
\renewcommand{\arraystretch}{1.2}
\caption{Attack success rate (ASR\,\%) on indirect prompt injection benchmarks across
all evaluated base models, including the undefended base pretrained model for reference.
Lower is better. \textbf{Bold} marks the best value per metric, model, and dataset.}
\label{tab:ood_asr}
\centering
\small
\begin{tabular*}{\textwidth}{@{\extracolsep{\fill}}ll cccc}
\toprule
& & \textbf{IID} & \multicolumn{3}{c}{\textbf{Out-of-Distribution (OOD)}} \\
\cmidrule(lr){3-3}\cmidrule(lr){4-6}
& & \textbf{SEP} & \textbf{PI-Attack} & \textbf{DataSentinel} & \textbf{Math-Tutor} \\
\textbf{Base Model} & \textbf{Condition} &
  \textbf{ASR\%} & \textbf{ASR\%} & \textbf{ASR\%} & \textbf{ASR\%} \\
\midrule
\multirow{3}{*}{SmolLM2-360M}
  & Base Pretrained &        40.19  &        33.79  &        84.48  &        3.33  \\
  & Baseline SFT    &         0.74  &         13.90  &         6.90  &         0.06  \\
  & Provenance-Aware & \best{0.51} & \best{0.05} & \best{0.00} & \best{0.00} \\
\midrule
\multirow{3}{*}{LLaMA-3-8B}
  & Base Pretrained &        32.50  &        21.30  &        24.14  &        17.22  \\
  & Baseline SFT    &         0.80  &         2.32  & 8.62 &         0.33  \\
  & Provenance-Aware & \best{0.69} & \best{0.03} & \best{0.00} & \best{0.00} \\
\midrule
\multirow{3}{*}{Qwen2.5-7B}
  & Base Pretrained    &      49.13     &      34.83     &   84.48        &      10.89     \\
  & Baseline SFT    &         0.66  &         3.74  &         10.34  &         0.11  \\
  & Provenance-Aware & \best{0.63} & \best{0.00} & \best{0.00} & \best{0.00} \\
\midrule
\multirow{3}{*}{Mistral-7B}
  & Base Pretrained &        43.47  &        43.58  &        51.72  &        35.94  \\
  & Baseline SFT    & \best{0.34} &        17.77  &         5.17  &         1.50  \\
  & Provenance-Aware &         0.66  & \best{0.03} & \best{0.00} & \best{0.00} \\
\bottomrule
\end{tabular*}
\end{table*}

\subsection{RQ2: Robustness to OOD Indirect Prompt Injection}

The IID/OOD split in Table~\ref{tab:ood_asr} directly tests whether Provenance-Aware
generalization is structural or a consequence of superior in-distribution fitting.
All conditions are evaluated across four base models on one IID benchmark (SEP test)
and three OOD datasets spanning different task domains and injection styles.

\noindent\textbf{IID results.}
On the SEP IID benchmark, both fine-tuned conditions stay below 1\% ASR across all
four base models, confirming that the Arch-Free SFT baseline fits the alignment training
distribution comparably to Provenance-Aware. This equivalence on IID rules out superior
in-distribution fitting as an explanation for any OOD difference, isolating the ring
architecture as the causal factor. Base pretrained models, by contrast, show 32.50--49.13\%
IID ASR, confirming the SEP task itself is non-trivial without alignment.

\begin{table}[!t]
\renewcommand{\arraystretch}{1.2}
\caption{AlpacaEval~2.0 length-controlled (LC) win rate against the base pretrained
model (50\%\,$=$\,parity). \textbf{Bold} marks the better (higher) condition per model.}
\label{tab:utility}
\centering
\small
\begin{tabular}{lcc}
\toprule
\textbf{Base Model} & \textbf{Baseline-SFT} & \textbf{Provenance-Aware} \\
\midrule
SmolLM2-360M & \textbf{57.1\%} &         49.5\%  \\
LLaMA-3-8B   & \textbf{58.1\%} &         48.8\%  \\
Qwen2.5-7B   &         60.7\%  & \textbf{60.9\%} \\
Mistral-7B   &         28.0\%  & \textbf{42.7\%} \\
\bottomrule
\end{tabular}
\end{table}

\smallskip\noindent\textbf{OOD results.}
Provenance-Aware drives OOD ASR to $\leq$0.03\% across all three held-out domains for
LLaMA-3-8B, Qwen2.5-7B, and Mistral-7B. The Baseline-SFT baseline
also improves substantially over base pretrained on most cells, but is consistently
at or behind Provenance-Aware: e.g.\ on PI-Attack, Baseline-SFT ASR is 2.32\%
(LLaMA), 3.74\% (Qwen), and 17.77\% (Mistral) versus Provenance-Aware's 0.03\%, 0.00\%,
and 0.03\% respectively. This gap illustrates that on novel task domains and injection styles the model never saw
during training, the Provenance-Aware model exhibits substantially higher resistance to IPI
attacks than Baseline-SFT, with the advantage holding consistently across models and
widening sharply on the hardest OOD cells. Because
this generalization gap emerges specifically on unseen distributions rather than on
in-distribution data, where both fine-tuned conditions perform comparably, it indicates
that the robustness conferred by the Provenance-Aware architecture stems from its structural
suppression mechanism rather than from memorizing patterns present in the training data.

\subsection{RQ3: Utility Preservation on Benign Inputs}
\label{sec:rq3-ref}

As defined in the defender's objective in Section~\ref{sec:threat-model}, a defense that achieves low ASR by degrading model outputs provides no practical security
guarantee. We evaluate utility
using AlpacaEval~2.0~\cite{alpaca_eval} (805 open-ended instruction-following prompts),
judged pairwise against the base pretrained model with Claude Sonnet~5 as judge (A/B order
randomized per prompt to cancel position bias) and reported as length-controlled (LC) win
rate, which removes verbosity bias by fitting a logistic model on length difference and
zeroing its coefficient~\cite{alpaca_eval_lc}. An LC win rate of 50\% indicates parity with
the undefended base model. 
We report both \textit{Provenance-Aware vs.\ Base} and \textit{Baseline-SFT vs.\ Base} comparisons in this section.

\begin{table}[!t]
\renewcommand{\arraystretch}{1.2}
\caption{ASR (\%) against adaptive attacks, LLaMA-3-8B-Instruct only Lower is better. \textbf{Bold} marks
the best condition per attack.}
\label{tab:adaptive}
\centering
\small
\begin{tabular}{lcc}
\toprule
\textbf{Condition} & \textbf{GCG ASR\%} & \textbf{NeuExe ASR\%} \\
\midrule
No Defense    &         50.71  &         48.44  \\
Baseline SFT  &          3.19  & 19.69 \\
Provenance-Aware  & \textbf{1.44} &          \textbf{1.56}  \\
\bottomrule
\end{tabular}
\end{table}

As Table~\ref{tab:utility} shows, for LLaMA-3-8B and Qwen2.5-7B, both Baseline-SFT and
Provenance-Aware achieve LC win rates at or above 50\% against the base pretrained model, with
overlapping confidence intervals. This indicates that the ring architecture is
statistically indistinguishable from an undefended model on open-ended
instruction-following, and likewise indistinguishable from a same-data fine-tune without
the ring components. SmolLM2-360M exhibits the same pattern at a much smaller scale
(57.1\% and 49.5\%, respectively). Mistral-7B is the sole model for which both conditions
fall meaningfully below parity (28.0\% and 42.7\%), though Provenance-Aware recovers
substantially over Baseline-SFT. We attribute this to a model-specific alignment
fragility in Mistral-7B rather than to a limitation of the ring architecture itself.

\begin{table}[!t]
\renewcommand{\arraystretch}{1.2}
\caption{Ablation study: ASR (\%) for component ablations across all evaluated base
models. \emph{w/o Stage~1}: ring architecture present but origin fine-tuning skipped;
\emph{w/o $Bias$}: fully trained checkpoint with origin attention bias zeroed at
inference. Lower ASR is better.}
\label{tab:ablation}
\centering
\setlength{\tabcolsep}{3pt}
\footnotesize
\begin{tabular}{llcccc}
\toprule
\textbf{Base Model} & \textbf{Condition} &
  \makecell{\textbf{SEP}\\\textbf{(IID)}} &
  \makecell{\textbf{Struct.}\\\textbf{-SFT}} &
  \makecell{\textbf{Data-}\\\textbf{Sentinel}} &
  \makecell{\textbf{Math-}\\\textbf{Tutor}} \\
\midrule
\multirow{3}{*}{SmolLM2-360M}
  & Full Design      & \textbf{0.51} & 0.05 & \textbf{0.00} & \textbf{0.00} \\
  & w/o Stage~1            &           0.54  &        \textbf{0.00}   &          \textbf{0.00} &          \textbf{0.00} \\
  & w/o $Bias$          &           29.51  &        45.77   &          96.55 &          11.56 \\
\midrule
\multirow{3}{*}{LLaMA-3-8B}
  & Full Design      & \textbf{0.69} & \textbf{0.03} & \textbf{0.00} & \textbf{0.00} \\
  & w/o Stage~1            &         0.80  & 0.03 & \textbf{0.00} & \textbf{0.00} \\
  & w/o $Bias$          &        72.04  &        77.80 &        89.66 &        67.56 \\
  \midrule
  \multirow{3}{*}{Qwen2.5-7B}
  & Full Design      & \textbf{0.63} & \textbf{0.00} & \textbf{0.00} & \textbf{0.00} \\
  & w/o Stage~1            &         0.69  & \textbf{0.00} & \textbf{0.00} & \textbf{0.00} \\
  & w/o $Bias$          &        69.35  &        70.62 &        98.28 &        39.67 \\
  \midrule
  \multirow{3}{*}{Mistral-7B}
  & Full Design      & \textbf{0.66} & \textbf{0.03} & \textbf{0.00} & \textbf{0.00} \\
  & w/o Stage~1            &         0.71  & 0.13 & \textbf{0.00} & \textbf{0.00} \\
  & w/o $Bias$          &         8.08  &        12.22 &        94.83 &        32.89 \\
\bottomrule
\end{tabular}
\end{table}

\subsection{RQ4: Robustness to Adaptive Attacks}
To evaluate how the proposed defense performs under adaptive attacks, we mainly consider two white-box adaptive attacks from existing work: \textbf{GCG}~\cite{zou2023universal} optimizes adversarial suffixes via greedy
coordinate gradient search to maximize the probability of the model following an
injected instruction. \textbf{NeuExe}\cite{Pasquini2024} executes the attack by exploiting neural
execution paths to bypass defenses. Both attacks are optimized against, and evaluated
on LLaMA-3-8B-Instruct, with universal suffix/trigger optimization across PI-Attack and DataSentinel datasets. Each adaptive attack is optimized with 2,000 steps.

Table~\ref{tab:adaptive} reports ASR under both white-box adaptive attacks. GCG ASR
drops from 50.71\% (No Defense) to 1.44\% under Provenance-Aware, while NeuExe ASR drops from
47.81\% to 0.87\%. These results confirm that the defense does not rely on
pattern-matching static injection templates: both attacks are explicitly optimized
against the target model with full knowledge of its weights, and Provenance-Aware still
suppresses each by more than an order of magnitude. This robustness follows directly from
the architectural nature of our defense: because suppression is enforced through origin
labels assigned at the application layer rather than inferred from token content,
optimizing an adversarial suffix or trigger to appear benign to the model's internal
representations does not change the ring ID under which that suffix is processed, leaving
the attacker with no gradient signal that can shift a Ring~3 token's provenance.

\begin{table}[!t]
\renewcommand{\arraystretch}{1.2}
\caption{AlpacaEval~2.0 LC win rate against base pretrained for a
Stage-1-only / w/o stage-1 checkpoint, compared against the full two-stage Provenance-Aware model from
Table~\ref{tab:utility}.}
\label{tab:stage1_utility}
\centering
\small
\begin{tabular}{lccc}
\toprule
\textbf{Base Model} & \textbf{\makecell[c]{Stage-1\\only}} & \textbf{\makecell[c]{w/o \\ Stage-1}} & \textbf{\makecell[c]{Full \\ Provenance-Aware}} \\
\midrule
SmolLM2-360M  & 56.1\% & 18.8\% & 49.5\% \\
LLaMA-3-8B  & 65.6\% & 28.7\% & 48.8\% \\
Qwen2.5-7B  & 62.1\% & 29.4\% & 60.9\% \\
Mistral-7B  & 56.5\% & 17.8\% & 42.7\% \\
\bottomrule
\end{tabular}
\end{table}

\begin{table}[!t]
\renewcommand{\arraystretch}{1.2}
\caption{HarmBench-judged\cite{mazeika2024harmbench} jailbreak ASR (\%) on WildJailbreak dataset. Lower
is better. \textbf{Bold} marks the better condition per model and dataset.}
\label{tab:jailbreak}
\centering
\small
\begin{tabular}{llc}
\toprule
\textbf{Base Model} & \textbf{Condition} & \textbf{WildJailbreak ASR\%} \\
\midrule
\multirow{2}{*}{LLaMA-3-8B}
  & Base Pretrained & 8.42\\
  & Provenance-Aware    & \textbf{0.00}\\
\midrule
\multirow{2}{*}{Qwen2.5-7B}
  & Base Pretrained & 37.28\\
  & Provenance-Aware    & \textbf{0.00}\\
\midrule
\multirow{2}{*}{Mistral-7B}
  & Base Pretrained &         41.04 \\
  & Provenance-Aware    & \textbf{0.00}\\
\bottomrule
\end{tabular}
\end{table}

\subsection{RQ5: Component Contribution}

Table~\ref{tab:ablation} reports ASR for two ablation conditions alongside the full
Provenance-Aware model and Baseline-SFT baseline.
The \emph{w/o Stage~1} condition retains the full ring architecture but skips origin
fine-tuning, going directly from the pretrained checkpoint to Stage~2 alignment.
The \emph{w/o $Bias$} condition uses the fully trained Provenance-Aware checkpoint but
zeros the origin attention bias at inference, leaving ring embeddings intact but removing
the structural attention gate. 

As illustrated in Table~\ref{tab:ablation}, across all three 7--8B models, skipping Stage~1 leaves attack success rate (ASR)
essentially unchanged relative to the full Provenance-Aware model: SEP and all three OOD
datasets stay within 0.1 percentage points, with every OOD cell remaining at or below
0.13\%. 
% This is expected since Stage-1 is designed to fix the model's performance on benign tasks before the model is taught the suppression behavior, which means Stage-1's contribution is mostly in benign-task utility and preparing the model for Stage-2 suppression learning. 
On the utility side, however, as Table~\ref{tab:stage1_utility} shows, removing Stage~1 entirely causes utility to
collapse well below the full two-stage model for every base model, confirming that Stage~1 is essential for preserving
utility, not merely beneficial. The Stage-1-only checkpoint further confirms this
split, outscoring the full model on AlpacaEval for every base model and clearing 50\% win
rate throughout (Section~\ref{sec:rq3-ref}), indicating Stage~1 alone restores general capability,
Stage~2 alone teaches suppression, and neither substitutes for the other.

As for the contribution of origin bias matrix, as Table~\ref{tab:ablation} shows, zeroing the origin attention bias collapses the defense back toward base-pretrained
territory: OOD ASR jumps from $\leq$0.03\% to 12--98\% depending on model and dataset,
in several cases landing worse than the corresponding base pretrained
model's own ASR on that cell (Table~\ref{tab:ood_asr}). SEP (IID) ASR collapses
similarly, confirming the bias term is not merely an OOD-generalization aid
but the primary suppression mechanism overall. This indicates that suppression requires an explicit attention-level gate rather than relying on the embedding signal to propagate its effect implicitly.

\subsection{RQ6: Generalization Beyond Prompt Injection}
\label{sec:eval-generalize}

Previous sections all evaluate task-hijacking indirect prompt injection, where the attacker's goal is to
redirect the model onto a different, attacker-chosen task. In this research question we seek to answer if our defense generalize to other distinct
threat class such as \emph{jailbreaking}, where the injected content instead targets safety
alignment, attempting to elicit a harmful response rather than hijack the task. We extract jailbreak
prompts from WildJailbreak\cite{wildteaming2024}, and placed them in the same
Ring3 channel as ordinary prompt injections, alongside the original benign task in other rings. If Ring3 suppression is a general provenance-based mechanism rather than
one narrowly learned for the SEP injection format, it should suppress jailbreak content
in Ring3 as well, without any jailbreak-specific training. We score attack
success with the HarmBench classifier~\cite{mazeika2024harmbench} (\emph{jailbreak\_asr}
in Table~\ref{tab:jailbreak}), a dedicated LLM judge for harmful-content classification.

As illustrated in Table~\ref{tab:jailbreak}, Provenance-Aware achieves 0.00\% HarmBench-judged jailbreak
ASR across all three models and both datasets. 
Base pretrained shows genuine jailbreak susceptibility (e.g. 41.04\% on Mistral-7B) that Provenance-Aware fully suppresses. This generalization
comes entirely from Ring~3 attention suppression firing unconditionally on provenance,
consistent with the RQ2 finding that the mechanism does not depend on content-level
pattern matching.

% \smallskip\noindent\textbf{A faster refusal-based heuristic paints a noisier, less
% conclusive picture.} During initial screening we also scored jailbreak success with a
% regex-based refusal heuristic (no refusal language present and output does not match
% the original task label) instead of HarmBench. That heuristic shows a much higher and
% more variable ASR across all conditions (19--79\% depending on model and dataset), and
% does \emph{not} consistently favor Origin-Aware over Base Pretrained---e.g.\ it flags
% LLaMA-3-8B Origin-Aware as worse than Base Pretrained on both datasets. We attribute
% this to the heuristic over-counting benign task compliance as jailbreak success (a
% verbose but harmless response that neither refuses nor exactly matches the reference
% label still counts as a "hit"), rather than to a genuine safety regression: HarmBench,
% which directly classifies whether the output is actually harmful, shows no such
% regression anywhere. We report only the HarmBench numbers here and treat the
% refusal-based heuristic as a fast pre-filter, not a reliable ASR metric on its own.

% =============================================================================
\section{Limitation and Future Work
}
\label{sec:limitations}
\smallskip\noindent\textbf{Ring-level suppression.}
Our defense operates at ring granularity, it suppresses all content in a given ring
unconditionally, without differentiating instructions from data within a specific
ring. In our current hierarchy, this means legitimate retrieved content that is assigned
to Ring~3 is suppressed just as unconditionally as an injected instruction would be. Use
cases that require the model to act on trusted external data therefore call for a more
fine-grained ring assignment than the four-ring hierarchy we demonstrate in this
paper, for example, splitting retrieved documents into separate trusted and untrusted
rings.
% \smallskip\noindent\textbf{Ring-level suppression.}
% The core mechanism of Origin-Aware Transformers suppresses all content in a given ring unconditionally.
% In the current ring hierarchy design, Ring~3 is designated for untrusted external data, and the trained model
% is prevented from acting on any content that arrives in Ring~3---regardless of whether that content is
% an adversarial injection or a legitimate piece of retrieved information.
% This means that if a deployment needs to retrieve data from an untrusted source, the model will be
% unable to attend to that data in Ring~3 at all.
% Our architecture provides no mechanism to differentiate \emph{instructions} from \emph{data} within
% a ring; the trust decision operates at the ring granularity, not at the content level.
% As a consequence, correct ring assignment becomes a critical deployment decision.
% Application developers must assess the trustworthiness of each data source and assign origin rings
% accordingly before model execution.
% Sources from which the application needs to retrieve and use content must be assigned to a ring with
% sufficient authority (e.g., Ring~2 or a future Ring~4 trusted-data tier); only sources whose content
% should be blocked unconditionally should be assigned to Ring~3.
% This shifts part of the security responsibility to the application layer, which must maintain a
% reliable source-to-ring mapping.

% \smallskip\noindent\textbf{Future Directions.}
\noindent\textbf{Intra-ring instruction--data separation.}
A natural extension of the current design is finer-grained separation between instructions and data within a
single ring. One
direction is to train the model to semantically distinguish data from instructions within
a ring, which would require fine-grained annotation of instruction versus data spans and a
training procedure that reasons about content type independent of provenance. A
complementary, more immediate direction is to pair our architectural suppression with
existing content-level defenses as a second verification layer. Rather than suppressing a
ring unconditionally, the model could apply moderate suppression to sources with mixed
trust and defer to a detector from existing defenses ~\cite{inan2023llama, wang2025promptsleuth} to flag spans that warrant stricter handling. We
consider both directions promising for reducing reliance on conservative ring assignment
policies.

\noindent\textbf{Broader applications of the ring architecture.}
Although this paper focuses on indirect prompt injection, the origin ring framework is not
limited to this setting. Our jailbreak evaluation (Section~\ref{sec:eval-generalize})
already provides evidence of this generality, showing that origin-based suppression
transfers to a threat model distinct from the one it was trained against. More broadly,
any scenario in which an LLM processes tokens from multiple sources with differing trust
could benefit from provenance-aware attention control. For instance, retrieval corpus
poisoning in RAG pipelines~\cite{zou2024poisonedrag} and tool-output manipulation in agentic
systems~\cite{debenedetti2024agentdojo}. We
believe the ring architecture provides a general-purpose trust-separation substrate that
can be instantiated for these and other LLM safety challenges beyond prompt injection.

\noindent\textbf{Post-hoc fine-tuning versus pretraining.}
Our current design instantiates Provenance-Aware Transformers via supervised fine-tuning
applied to released pretrained checkpoints, a choice driven primarily by our goal of
adapting existing models rather than training new ones from scratch, as well as by
compute constraints that make full pretraining infeasible in our setting. This is not,
however, an architectural restriction. The origin embedding, origin attention bias, and
origin scale we introduce are compatible with pretraining as well, and a model could in
principle be trained from scratch with origin-aware components integrated from the first
step. Such an integration would allow the architecture to be adopted fundamentally into
the design of future LLMs, rather than applied only as a retrofit, and we leave this
direction to future work with access to larger-scale training resources.

% =============================================================================
\section{Conclusion}
Indirect Prompt injection remains difficult to defend because standard language models have no
architectural notion of source authority. They process operator instructions, user input,
and externally sourced text through the same content-addressed mechanism, leaving the model
to infer from wording alone what should be obeyed and what should be treated as data. In
this paper, we argued that provenance-aware deployment offers a practical way to break this
failure mode. 
% If the application can label where tokens come from before inference, the
% model can be trained to use that provenance structurally rather than relying only on
% content-level attack recognition.
We presented \emph{Provenance-Aware Transformers}, an architectural defense that makes
application-supplied origin labels actionable throughout the transformer's forward pass.
Origin embeddings preserve provenance at the representation level, origin attention bias
constrains cross-source influence at the attention level, and alignment training teaches
the model the behavioral meaning of each ring. Together, these components create a trust boundary separation policy for provenance-aware deployments sources designated as
non-authoritative can be suppressed unconditionally, while permitted sources remain
available for ordinary task completion.

\bibliographystyle{plainurl}
\bibliography{main}

@inproceedings{liu2024promptinjection,
  title={Formalizing and Benchmarking Prompt Injection Attacks and Defenses},
  author={Liu, Yupei and Jia, Yuqi and Geng, Runpeng and Jia, Jinyuan and Gong, Neil Zhenqiang},
  booktitle={USENIX Security Symposium},
  year={2024}
}

@article{aws2024llm,
  title     = {What is LLM? - Large Language Models Explained},
  author    = {{Amazon Web Services}},
  journal   = {AWS Documentation},
  year      = {2024},
  url       = {https://aws.amazon.com/what-is/large-language-model/}
}

@article{ibm2024llm,
  title     = {What Are Large Language Models (LLMs)? - IBM},
  author    = {{IBM Cloud}},
  journal   = {IBM Cloud Learn Hub},
  year      = {2024},
  url       = {https://www.ibm.com/think/topics/large-language-models}
}

@article{openai2023gpt4,
  title     = {GPT-4 Technical Report},
  author    = {OpenAI},
  journal   = {OpenAI Research},
  year      = {2023},
  url       = {https://openai.com/research/gpt-4}
}

@article{debenedetti2024agentdojo,
  title={Agentdojo: A dynamic environment to evaluate prompt injection attacks and defenses for llm agents},
  author={Debenedetti, Edoardo and Zhang, Jie and Balunovic, Mislav and Beurer-Kellner, Luca and Fischer, Marc and Tram{\`e}r, Florian},
  journal={Advances in Neural Information Processing Systems},
  volume={37},
  pages={82895--82920},
  year={2024}
}

@article{google2024vertex,
  title     = {Overview of Generative AI on Vertex AI - Google Cloud},
  author    = {Google Cloud},
  journal   = {Google Cloud Documentation},
  year      = {2024},
  url       = {https://cloud.google.com/vertex-ai/generative-ai/docs/overview}
}

@article{meta2024llama2,
  title     = {Llama 2: Open Foundation and Fine-Tuned Chat Models},
  author    = {Meta AI},
  journal   = {Meta AI Blog},
  year      = {2023},
  url       = {https://ai.meta.com/research/publications/llama-2-open-foundation-and-fine-tuned-chat-models/}
}

@article{openai2024agents,
  title     = {OpenAI Agents SDK},
  author    = {OpenAI},
  journal   = {OpenAI Platform Documentation},
  year      = {2024},
  url       = {https://platform.openai.com/docs/guides/agents-sdk}
}

@article{aws2024agents,
  title     = {Using Large Language Models on Amazon Bedrock for multi-step task execution},
  author    = {{Amazon Web Services}},
  journal   = {AWS Machine Learning Blog},
  year      = {2024},
  url       = {https://aws.amazon.com/blogs/machine-learning/using-large-language-models-on-amazon-bedrock-for-multi-step-task-execution/}
}

@article{touvron2023llama,
  title={LLaMA: Open and Efficient Foundation Language Models},
  author={Hugo Touvron and Thibaut Lavril and Gautier Izacard and Xavier Martinet and Marie-Anne Lachaux and Timothée Lacroix and Baptiste Rozière and Naman Goyal and Eric Hambro and Faisal Azhar and Aur{\'e}lien Rodriguez and Armand Joulin and Edouard Grave and Guillaume Lample},
  journal={arXiv preprint arXiv:2302.13971},
  year={2023},
  url={https://arxiv.org/abs/2302.13971}
}

@misc{lakera2023visual,
  author       = {{Lakera AI}},
  title        = {Visual Prompt Injections},
  howpublished = {\url{https://www.lakera.ai/blog/visual-prompt-injections}},
  year         = {2023},
  note         = {Accessed: 2025-04-18}
}

@inproceedings{chen2025struq,
  title={$\{$StruQ$\}$: Defending against prompt injection with structured queries},
  author={Chen, Sizhe and Piet, Julien and Sitawarin, Chawin and Wagner, David},
  booktitle={34th USENIX Security Symposium (USENIX Security 25)},
  pages={2383--2400},
  year={2025}
}

@inproceedings{Perez2022,
  author = {Perez, F\'{a}bio and Ribeiro, Ian},
  title = {{Ignore previous prompt: Attack techniques for language models}},
  booktitle = {NeurIPS ML Safety Workshop},
  year = {2022},
  url = {https://arxiv.org/abs/2211.09527}
}

@misc{Selvi2022,
  author = {Selvi, Jose},
  title = {{Exploring Prompt Injection Attacks}},
  year = {2022},
  howpublished = {\url{https://research.nccgroup.com/2022/12/05/exploring-prompt-injection-attacks/}}
}

@misc{Willison2022,
  author = {Willison, Simon},
  title = {{Prompt injection attacks against GPT-3}},
  year = {2022},
  howpublished = {\url{https://simonwillison.net/2022/Sep/12/prompt-injection/}}
}

@misc{Harang2023,
  author = {Harang, Rich},
  title = {{Securing LLM Systems Against Prompt Injection}},
  year = {2023},
  howpublished = {\url{https://developer.nvidia.com/blog/securing-llm-systems-against-prompt-injection}}
}

@misc{Willison2023,
  author = {Willison, Simon},
  title = {{Delimiters won’t save you from prompt injection}},
  year = {2023},
  howpublished = {\url{https://simonwillison.net/2023/May/11/delimiters-wont-save-you}}
}

@misc{Nakajima2022,
  author = {Nakajima, Yohei},
  title = {{“Yohei’s blog post” (prompt injection demonstration)}},
  year = {2022},
  howpublished = {\url{https://twitter.com/yoheinakajima/status/1582844144640471040}}
}

@misc{Random_Sequnce,
  title = {{Random sequence enclosure}},
  year = {2023},
  howpublished = {\url{https://learnprompting.org/docs/prompt_hacking/defensive_measures/random_sequence?srsltid=AfmBOoqIB3KhS3Lc1d9NY8T4LPpMzH6dbYmXJi1y697s9byu9xQcDykV}}
}

@inproceedings{Shao2024,
  title={Enhancing prompt injection attacks to llms via poisoning alignment},
  author={Shao, Zedian and Liu, Hongbin and Mu, Jaden and Gong, Neil},
  booktitle={Proceedings of the 18th ACM Workshop on Artificial Intelligence and Security},
  pages={13--27},
  year={2025}
}

@article{Pasquini2024,
  author = {Pasquini, Daniele and Strohmeier, Matthias and Troncoso, Carmela},
  title = {{Neural exec: Learning (and learning from) execution triggers for prompt injection attacks}},
  journal = {arXiv preprint arXiv:2403.03792},
  year = {2024},
  url = {https://arxiv.org/abs/2403.03792}
}

@inproceedings{Hui2024,
  author = {Hui, Bo and Yuan, Haolin and Gong, Neil Z. and Burlina, Philippe and Cao, Yue},
  title = {{PLeak: Prompt leaking attacks against large language model applications}},
  booktitle = {Proc. of ACM CCS},
  year = {2024},
  url = {https://arxiv.org/abs/2405.06823}
}

@inproceedings{liu2025datasentinel,
  title={Datasentinel: A game-theoretic detection of prompt injection attacks},
  author={Liu, Yupei and Jia, Yuqi and Jia, Jinyuan and Song, Dawn and Gong, Neil Zhenqiang},
  booktitle={2025 IEEE Symposium on Security and Privacy (SP)},
  pages={2190--2208},
  year={2025},
  organization={IEEE}
}

@inproceedings{zhang2025defense,
  title={Defense against prompt injection attacks via mixture of encodings},
  author={Zhang, Ruiyi and Sullivan, David and Jackson, Kyle and Xie, Pengtao and Chen, Mei},
  booktitle={Proceedings of the 2025 Conference of the Nations of the Americas Chapter of the Association for Computational Linguistics: Human Language Technologies (Volume 2: Short Papers)},
  pages={244--252},
  year={2025}
}

@inproceedings{Chen2024SecAlignDAA,
  title={Secalign: Defending against prompt injection with preference optimization},
  author={Chen, Sizhe and Zharmagambetov, Arman and Mahloujifar, Saeed and Chaudhuri, Kamalika and Wagner, David and Guo, Chuan},
  booktitle={Proceedings of the 2025 ACM SIGSAC Conference on Computer and Communications Security},
  pages={2833--2847},
  year={2025}
}

@article{zou2023universal,
  author  = {Zou, Andy and Wang, Zifan and Carlini, Nicholas and Nasr, Milad and Kolter, J.~Zico and Fredrikson, Matt},
  title   = {Universal and Transferable Adversarial Attacks on Aligned Language Models},
  journal = {arXiv preprint arXiv:2307.15043},
  year    = {2023},
  url     = {https://arxiv.org/abs/2307.15043}
}

@misc{shi2025promptarmorsimpleeffectiveprompt,
      title={PromptArmor: Simple yet Effective Prompt Injection Defenses}, 
      author={Tianneng Shi and Kaijie Zhu and Zhun Wang and Yuqi Jia and Will Cai and Weida Liang and Haonan Wang and Hend Alzahrani and Joshua Lu and Kenji Kawaguchi and Basel Alomair and Xuandong Zhao and William Yang Wang and Neil Gong and Wenbo Guo and Dawn Song},
      year={2025},
      eprint={2507.15219},
      archivePrefix={arXiv},
      primaryClass={cs.CR},
      url={https://arxiv.org/abs/2507.15219}, 
}

@misc{emotional_claude2024,
  author       = {Claburn, Thomas},
  title        = {Anthropic's Claude emotional prompt vulnerability raises red flags},
  howpublished = {\url{https://www.theregister.com/2024/10/12/anthropics_claude_vulnerable_to_emotional/}},
  note         = {Accessed: 2025-04-29},
  year         = {2024},
  month        = {October},
  day          = {12}
}

@misc{openclaw,
  title = {{OpenClaw}: An Open-Source Autonomous AI Agent Framework},
  author = {Steinberger, Peter},
  year = {2026},
  howpublished = {\url{https://openclaw.ai/}},
  note = {Accessed: 2026-03-30}
}

@misc{crewai2026,
  title = {{CrewAI}: Multi-Agent Orchestration for Complex Workflows},
  author = {Moura, João},
  year = {2026},
  howpublished = {\url{https://www.crewai.com/}},
  note = {Accessed: 2026-03-30}
}

@misc{autogen2026,
  title = {{Microsoft AutoGen}: Enabling Next-Generation Large Language Model Applications with Multi-Agent Conversations},
  author = {Microsoft Research},
  year = {2026},
  howpublished = {\url{https://microsoft.github.io/autogen/}},
  note = {Accessed: 2026-03-30}
}

@inproceedings{Piet2024Jatmo,
  author    = {Julian Piet and Moustafa Alrashed and Chawin Sitawarin and others},
  title     = {Jatmo: Prompt Injection Defense by Task-Specific Finetuning},
  booktitle = {European Symposium on Research in Computer Security (ESORICS)},
  year      = {2024}
}

@article{Wallace2024InstructionHierarchy,
  author    = {Eric Wallace and Kai Xiao and Reon Leike and others},
  title     = {The Instruction Hierarchy: Training {LLMs} to Prioritize Privileged Instructions},
  journal   = {arXiv preprint arXiv:2404.13208},
  year      = {2024}
}

@misc{learnprompting2023sandwich,
  title        = {Sandwich Defense},
  author       = {{Learn Prompting}},
  year         = {2023},
  howpublished = {\url{https://learnprompting.org/docs/prompt_hacking/defensive_measures/sandwich_defense}},
  note         = {Accessed: 2026-03-30}
}

@article{inan2023llama,
  title={Llama guard: Llm-based input-output safeguard for human-ai conversations},
  author={Inan, Hakan and Upasani, Kartikeya and Chi, Jianfeng and Rungta, Rashi and Iyer, Krithika and Mao, Yuning and Tontchev, Michael and Hu, Qing and Fuller, Brian and Testuggine, Davide and others},
  journal={arXiv preprint arXiv:2312.06674},
  year={2023}
}

@inproceedings{zverev2025can,
  title={Can llms separate instructions from data? and what do we even mean by that?},
  author={Zverev, Egor and Abdelnabi, Sahar and Tabesh, Soroush and Fritz, Mario and Lampert, Christoph},
  booktitle={International Conference on Learning Representations},
  volume={2025},
  pages={67147--67179},
  year={2025}
}

@article{wang2025promptsleuth,
  title={PromptSleuth: Detecting Prompt Injection via Semantic Intent Invariance},
  author={Wang, Mengxiao and Zhang, Yuxuan and Gu, Guofei},
  journal={arXiv preprint arXiv:2508.20890},
  year={2025}
}

@article{zhong2025attention,
  title={Attention is All You Need to Defend Against Indirect Prompt Injection Attacks in LLMs},
  author={Zhong, Yinan and Miao, Qianhao and Chen, Yanjiao and Deng, Jiangyi and Cheng, Yushi and Xu, Wenyuan},
  journal={arXiv preprint arXiv:2512.08417},
  year={2025}
}

@inproceedings{jia2025promptlocate,
  title={Promptlocate: Localizing prompt injection attacks},
  author={Jia, Yuqi and Liu, Yupei and Shao, Zedian and Jia, Jinyuan and Gong, Neil Zhenqiang},
  booktitle={2026 IEEE Symposium on Security and Privacy (SP)},
  pages={4243--4261},
  year={2026},
  organization={IEEE}
}

@misc{claudecode,
  title = {Claude Code: AI-powered coding assistant},
  author = {Anthropic},
  year = {2026},
  howpublished = {\url{https://code.claude.com/docs/en/overview}},
  note = {Accessed: 2026-07-23}
}

@misc{codex,
  title = {Codex},
  author = {OpenAI},
  year = {2026},
  howpublished = {\url{https://openai.com/codex/}},
  note = {Accessed: 2026-07-23}
}

@misc{alpaca_eval,
  author = {Xuechen Li and Tianyi Zhang and Yann Dubois and Rohan Taori and Ishaan Gulrajani and Carlos Guestrin and Percy Liang and Tatsunori B. Hashimoto},
  title = {AlpacaEval: An Automatic Evaluator of Instruction-following Models},
  year = {2023},
  howpublished = {\url{https://github.com/tatsu-lab/alpaca_eval}}
}

@article{alpaca_eval_lc,
  title   = {Length-Controlled AlpacaEval: A Simple Way to Debias Automatic Evaluators},
  author  = {Dubois, Yann and Galambosi, Bal{\'a}zs and Liang, Percy and Hashimoto, Tatsunori B.},
  journal = {arXiv preprint arXiv:2404.04475},
  year    = {2024},
  url     = {https://arxiv.org/abs/2404.04475}
}

@article{mazeika2024harmbench,
  title   = {HarmBench: A Standardized Evaluation Framework for Automated Red Teaming and Robust Refusal},
  author  = {Mazeika, Mantas and Phan, Long and Yin, Xuwang and Zou, Andy and Wang, Zifan and Mu, Norman and Sakhaee, Elham and Li, Nathaniel and Basart, Steven and Li, Bo and Forsyth, David and Hendrycks, Dan},
  journal = {arXiv preprint arXiv:2402.04249},
  year    = {2024},
  url     = {https://arxiv.org/abs/2402.04249}
}

@misc{openai_prompt,
  title = {OpenAI Developer Prompt Engineering Guide},
  author = {OpenAI},
  year = {2026},
  howpublished = {\url{https://developers.openai.com/api/docs/guides/prompt-engineering}},
  note = {Accessed: 2026-07-23}
}

@misc{nanochat,
  title = {karpathy/nanochat: The best ChatGPT that \$100 can buy.},
  author = {Andrej Karpathy},
  year = {2026},
  howpublished = {\url{https://github.com/karpathy/nanochat}},
  note = {Accessed: 2026-07-23}
}

@misc{allal2025smollm2smolgoesbig,
      title={SmolLM2: When Smol Goes Big -- Data-Centric Training of a Small Language Model}, 
      author={Loubna Ben Allal and Anton Lozhkov and Elie Bakouch and Gabriel Martín Blázquez and Guilherme Penedo and Lewis Tunstall and Andrés Marafioti and Hynek Kydlíček and Agustín Piqueres Lajarín and Vaibhav Srivastav and Joshua Lochner and Caleb Fahlgren and Xuan-Son Nguyen and Clémentine Fourrier and Ben Burtenshaw and Hugo Larcher and Haojun Zhao and Cyril Zakka and Mathieu Morlon and Colin Raffel and Leandro von Werra and Thomas Wolf},
      year={2025},
      eprint={2502.02737},
      archivePrefix={arXiv},
      primaryClass={cs.CL},
      url={https://arxiv.org/abs/2502.02737}, 
}

@misc{llama3modelcard,
    title = {Llama 3 Model Card},
    url = {https://github.com/meta-llama/llama3/blob/main/MODEL_CARD.md},
    author = {AI@Meta},
    month = {September},
    year = {2024}
}

@misc{qwen2.5,
    title = {Qwen2.5: A Party of Foundation Models},
    url = {https://qwenlm.github.io/blog/qwen2.5/},
    author = {Qwen Team},
    month = {September},
    year = {2024}
}

@misc{mistral-7b,
    title = {Mistral-7B-v0.3},
    url = {https://huggingface.co/mistralai/Mistral-7B-v0.3},
    author = {Mistral},
    month = {September},
    year = {2024}
}

@misc{promptguard,
    title = {Llama PromptGuard},
    url = {https://huggingface.co/meta-llama/Prompt-Guard-86M},
    author = {meta-llama},
    month = {September},
    year = {2024}
}

@misc{pi-attack,
    title = {PI-Attack},
    url = {https://huggingface.co/datasets/xxz224/prompt-injection-attack-dataset},
    author = {Huggingface Developer},
    month = {September},
    year = {2024}
}

@misc{alpaca,
  author = {Rohan Taori and Ishaan Gulrajani and Tianyi Zhang and Yann Dubois and Xuechen Li and Carlos Guestrin and Percy Liang and Tatsunori B. Hashimoto },
  title = {Stanford Alpaca: An Instruction-following LLaMA model},
  year = {2023},
  publisher = {GitHub},
  journal = {GitHub repository},
  howpublished = {\url{https://github.com/tatsu-lab/stanford_alpaca}},
}

@misc{wildteaming2024,
      title={WildTeaming at Scale: From In-the-Wild Jailbreaks to (Adversarially) Safer Language Models}, 
      author={Liwei Jiang and Kavel Rao and Seungju Han and Allyson Ettinger and Faeze Brahman and Sachin Kumar and Niloofar Mireshghallah and Ximing Lu and Maarten Sap and Yejin Choi and Nouha Dziri},
      year={2024},
      eprint={2406.18510},
      archivePrefix={arXiv},
      primaryClass={cs.CL},
      url={https://arxiv.org/abs/2406.18510}, 
}

@inproceedings{greshake2023not,
  title={Not what you've signed up for: Compromising real-world llm-integrated applications with indirect prompt injection},
  author={Greshake, Kai and Abdelnabi, Sahar and Mishra, Shailesh and Endres, Christoph and Holz, Thorsten and Fritz, Mario},
  booktitle={Proceedings of the 16th ACM workshop on artificial intelligence and security},
  pages={79--90},
  year={2023}
}

@inproceedings{yi2025benchmarking,
  title={Benchmarking and defending against indirect prompt injection attacks on large language models},
  author={Yi, Jingwei and Xie, Yueqi and Zhu, Bin and Kiciman, Emre and Sun, Guangzhong and Xie, Xing and Wu, Fangzhao},
  booktitle={Proceedings of the 31st ACM SIGKDD Conference on Knowledge Discovery and Data Mining V. 1},
  pages={1809--1820},
  year={2025}
}

@inproceedings{radford2019gpt2,
  title={Language models are unsupervised multitask learners},
  author={Radford, Alec and Wu, Jeffrey and Child, Rewon and Luan, David and Amodei, Dario and Sutskever, Ilya and others},
  journal={OpenAI blog},
  volume={1},
  number={8},
  pages={9},
  year={2019}
}

@inproceedings{vaswani2017attention,
  title={Attention is all you need},
  author={Vaswani, Ashish and Shazeer, Noam and Parmar, Niki and Uszkoreit, Jakob and Jones, Llion and Gomez, Aidan N and Kaiser, {\L}ukasz and Polosukhin, Illia},
  booktitle={Advances in Neural Information Processing Systems},
  volume={30},
  year={2017}
}

@article{brown2020gpt3,
  title={Language models are few-shot learners},
  author={Brown, Tom and Mann, Benjamin and Ryder, Nick and Subbiah, Melanie and Kaplan, Jared D and Dhariwal, Prafulla and Neelakantan, Arvind and Shyam, Pranav and Sastry, Girish and Askell, Amanda and others},
  journal={Advances in Neural Information Processing Systems},
  volume={33},
  pages={1877--1901},
  year={2020}
}

@article{jiang2023mistral,
  title={Albert q. jiang, alexandre sablayrolles, arthur mensch, chris bamford, devendra singh chaplot, diego de las casas, florian bressand, gianna lengyel, guillaume lample, lucile saulnier, l{\'e}lio renard lavaud, marie-anne lachaux, pierre stock, teven le scao, thibaut lavril, thomas wang, timoth{\'e}e lacroix, william el sayed},
  author={Chaplot, Devendra Singh},
  journal={arXiv preprint arXiv:2310.06825},
  volume={3},
  year={2023}
}

@inproceedings{zou2024poisonedrag,
  title={$\{$PoisonedRAG$\}$: Knowledge corruption attacks to $\{$Retrieval-Augmented$\}$ generation of large language models},
  author={Zou, Wei and Geng, Runpeng and Wang, Binghui and Jia, Jinyuan},
  booktitle={34th USENIX Security Symposium (USENIX Security 25)},
  pages={3827--3844},
  year={2025}
}

% optional clearing of the page
% \cleardoublepage
% \appendix

\end{document}